\documentclass[a4paper,11pt]{article}
\usepackage{jcappub} % for details on the use of the package, please see the JINST-author-manual
\usepackage{lineno}
\usepackage{amscd}
\usepackage{amsmath}
\usepackage{amssymb}
\usepackage{amsthm}
\newcommand{\Lya}{Ly$\alpha$ }
\newcommand{\LyaNo}{Ly$\alpha$}
\newcommand{\MathLya}{Ly\alpha}
\newcommand{\SFR}{\dot{M_*}}
\newcommand{\SFRD}{\dot{\rho_*}}

\title{Radiative Transfer Simulations of \Lya Intensity Mapping During Cosmic Reionization Including Sources from Galaxies and the Intergalactic Medium}

\author[a]{Abigail E. Ambrose}
\affiliation[a]{Ritter Astrophysical Research Center, University of Toledo, Dept.~of Physics and Astronomy, 2801  West Bancroft Street Toledo, Ohio 43606, USA}

\author[a]{Eli Visbal}
\author[a,b]{Mihir Kulkarni}
\affiliation[b]{Institut für Astrophysik und Geophysik, Georg-August Universität Göttingen, Friedrich-Hund-Platz 1, D-37077 Göttingen, Germany}

\author[c]{Matthew McQuinn}
\affiliation[c]{University of Washington, Department of Astronomy, 3910 15th Avenue NE, Seattle, WA 98195, USA}

\emailAdd{abigail.ambrose@rockets.utoledo.edu, elijah.visbal@utoledo.edu, mihir.kulkarni@uni-goettingen.de, mcquinn@uw.edu}

\abstract{We present new simulations of Lyman-$\alpha$ (\LyaNo) intensity maps that include \Lya radiative transfer in the intergalactic medium (IGM) and all significant sources of \Lya photons. The sources considered include \Lya directly from galaxies, cooling at the edges of ionized bubbles, recombinations within these bubbles, and reprocessing of galaxy continuum emission in the IGM. We also vary astrophysical parameters including the average neutral fraction of the IGM, the dust absorption of \Lya in galaxies, and the ionizing escape fraction. Previous work has suggested that \Lya intensity mapping can be used to constrain the neutral fraction of the IGM when accounting for radiative transfer in the IGM. When radiative transfer is ignored, direct \Lya emission from galaxies has the highest amplitude of power on all scales. When we include radiative transfer in our simulations, we find continuum emission reprocessed as \Lya is comparable to the \Lya emission directly from galaxies on large scales. For high neutral fraction in the IGM, emission from recombinations is comparable to galaxies on large scales. We find that the slope of the power spectrum is sensitive to the neutral fraction of the IGM when radiative transfer is included, suggesting that this may be useful for placing constraints on cosmic reionization. In addition, we find the power of galaxies is decreased across all scales due to dust absorption. We also find the escape fraction must be large for recombinations and bubble edges to contribute significantly to the power. We find the cross power is observable between SPHEREx \Lya intensity maps and a hypothetical galaxy survey is observable with a total signal-to-noise of 4 from $k=0.035$~Mpc$^{-1}$ to $k=1$~Mpc$^{-1}$. }

\begin{document}
\maketitle
\flushbottom

\section{Introduction} \label{sec:intro}

The Epoch of Reionization (EoR) is an exciting frontier of research in cosmology. There have been many observations to constrain the properties of reionization. One of these observations is the damping wing observed in the spectrum of galaxies and quasars, indicating that EoR ended as late as a redshift $z\approx5.3$. \citep{Iye2006,Ono2012,Schenker2012,Zhu2024}. The Cosmic Microwave Background (CMB) indicates that the Universe was less than 10\% reionized for redshifts $z>10$ and that the midpoint of reionization occurred between a redshift of $z=7.8$ and $z=8.8$. Additionally, kinetic Sunyaev-Zeldovich (kSZ) effect measurements constrain the length of reionization to $\Delta z<3$ \citep{Planck2016}. Upcoming observations will allow us to tighten these constraints and gain a better understanding of reionization history and the sources driving it (see \cite{Wise2019,Robertson2022} for recent reviews). 

Intensity mapping is a promising new technique that could be used to probe cosmic reionization. This technique involves creating large-scale 3-dimensional (3D) maps of spectral line emission from galaxies and/or the intergalactic medium (IGM,~\cite{Visbal2010,Bernal2022}). A key feature of these maps is they measure the combined emission from all sources including faint sources that are typically below the detection limit of traditional galaxy redshift surveys. A number of experiments have been proposed targeting a variety of lines such as Hydrogren Epoch of Reionization Array (HERA) mapping 21~cm emission, CO Mapping Array Project (COMAP) mapping CO emission, and CarbON CII line in post-rEionization and ReionizaTiOn epoch (CONCERTO) mapping CII emission \citep{DeBoer2017,Cleary2022,Dumitru2019}. 

Future telescopes such as Spectro-Photometer for the History of the Universe, Epoch of Reionization and Ices Explorer (SPHEREx) and the proposed Cosmic Dawn Intensity Mapper (CDIM) will be able to observe Lyman-$\alpha$ (\LyaNo) during reionization~\citep{Cheng2022,Alibay2023,Cooray2019}. \Lya will be a powerful line for this time period since it is a hydrogen recombination line, so no metals are necessary for these detections. The luminosity of \Lya is also dependent on the star formation rate (SFR) making it an excellent tracer for the cosmic star formation history (e.g. \cite{Silva2013,Pullen2014,Heneka2021,Padmanabhan2024}).

The authors of \cite{Visbal2018} suggested that \Lya intensity mapping would also constrain the timing of reionization because \Lya photons are easily scattered by the neutral IGM; however such radiative transfer was not included in previous models. This study showed that the \Lya power spectrum is sensitive to the global ionization fraction, so small-scale power was suppressed for an increasing neutral fraction due to the smoothing of the signal from intergalactic scattering off neutral regions. However, this work only focused on the signal from the galaxies and did not include other sources of \Lya photons.

In this paper, we model all significant sources of \Lya photons, including from galaxies, the edges of ionized bubbles, recombinations in the IGM, and reprocessing of galaxy continuum emission. We span a large parameter space in these simulations, including three different neutral fractions, four different values for dust absorption in galaxies, and three values for the escape fraction of ionizing photons. Finally, we forecast the cross-power from SPHEREx \Lya intensity maps and a hypothetical galaxy survey in order to assess the detectability. 

This paper is structured as follows. In Section~\ref{sec:method} we discuss our simulations and the implementation of each source. In Section~\ref{sec:results} we present our intensity maps, power spectra, and how different parameters effect these. Finally, in Section~\ref{sec:conclusions} we conclude with discussion of the implications of this work. Throughout this paper we assume $\Lambda$CDM with values $\Omega_\mathrm{m}=0.32$, $\Omega_\Lambda=0.68$, $\Omega_\mathrm{b}=0.049$, and $h=0.67$, consistent with the analysis in \cite{Planck2018}. All cosmological distances are given in comoving units unless otherwise specified.

\section{Methodology} \label{sec:method}

In this section, we describe our methods for generating \Lya intensity maps. We utilize 21cmFAST \citep{Mesinger2011}, details of which can be found in Section~\ref{21cm}. Then, we describe our modeling of the Ly$\alpha$ emission from all sources, including \Lya photons from galaxies, edges of ionized bubbles, recombinations in the IGM, and ultraviolet continuum emission from galaxies cascading down to Ly$\alpha$ photons in Sections \ref{Halos}-\ref{Continuum}. We describe our \Lya radiative transfer code in Section~\ref{RadTrans}. 

\subsection{21cmFAST Simulations} \label{21cm}

We use the semi-numeric simulation 21cmFAST \citep{Mesinger2011} to simulate the IGM density, the IGM neutral fraction, velocity field, and the halo density field. For the IGM density, initial conditions for the density and velocity are set and then the gravitational collapse is evolved according to first-order Lagrangian perturbation theory. Similarly, the velocity field is also given initial conditions and evolved following first-order perturbation theory. For the IGM neutral fraction, the fast Fourier radiative transfer (FFRT) method presented in \cite{Zahn2007} with an excursion-set approach is used to generate maps of HII regions using the density field. This method has been shown to reproduce the HII region distribution in full radiative transfer simulations of reionization \citep{Zahn2007}. The halo density field is created by finding overdensities in the IGM density that exceed the barrier for collapse in Sheth-Tormen \citep{Sheth1999,Mesinger2011}. 
We set the minimum halo mass that hosts a galaxy to be $1.5\times10^9~M_{\odot}$ based on the analytic work of~\cite{Furlanetto2017}. This analytic work as well as galaxy simulations, such as \cite{yeh2023}, shows that sources down to $M_{\mathrm{min}}\approx 10^8~M_{\odot}$ can host star formation, but the efficiency is reduced by feedback below $10^9~M_{\odot}$. 

We run these simulations for different values of cosmic neutral fraction and the ionizing escape fraction from galaxies. Our simulations are at a redshift of $z=7$ with a size of $(200~\text{Mpc})^3$ and a resolution of $256^3$ voxels. Two astrophysical parameters in 21cmFAST also affect the properties of reionization including the maximum radius of an ionized region $R_{\mathrm{max}}$, and ionizing efficiency $\zeta$. For all runs we use $R_{\mathrm{max}}=50$~Mpc, however, we note larger values would yield similar reionization morphology results. The ionizing efficiency is calculated as $\zeta=N_{\gamma}f_\mathrm{esc}f_{\ast}$ where $N_{\gamma}$ is the number of ionizing photons per stellar baryon,  $f_\mathrm{esc}$ is the escape fraction of ionizing radiation as a function of halo mass, and $f_*$ is the star formation efficiency as a function of halo mass. We will discuss how the escape fraction and star formation efficiency are determined in the following section.

\subsection{Ly$\alpha$ Photons from Galaxies}\label{Halos}

For each galaxy in the simulation box, we assume the \Lya luminosity is proportional to the SFR similar to \cite{Visbal2018}. In addition, we include fraction of \Lya absorbed by dust, $f_{\mathrm{dust}}$, use the star formation rate prescription from 21cmFAST, and allow the escape fraction of ionizing radiation, $f_{\mathrm{esc}}$, to vary with halo mass. This results in a total \Lya luminosity from a galaxy of
\begin{equation}
    L_{\mathrm{gal}}=2.0\times 10^{42}(1-f_{\mathrm{esc}})(1-f_{\mathrm{dust}})\frac{\dot{M_*}}{\text{M}_{\odot}\text{yr}^{-1}}\text{erg s}^{-1},
\end{equation}
where $\dot{M_*}$ is the SFR of the galaxy. This equation assumes each ionization results in $0.6$~\Lya photons and a Salpeter initial mass function \citep{Schaerer2003}.

Dust absorbs Ly$\alpha$ photons.\footnote{ We assume that dust only absorbs \Lya photons and not continuum photons -- while this is an extreme assumption, it is likely that the destruction of \Lya photons by dust is much larger because they scatter around the galaxy \citep{Pullen2014}.}  We allow the dust absorption fraction, $f_{\mathrm{dust}}$, to vary as 
\begin{equation}
    f_{\mathrm{dust}}=1-C_{\mathrm{dust}}\times 10^{-3}(1+z)^{\epsilon_{\mathrm{dust}}},
\end{equation}
where $C_{\mathrm{dust}}$ and $\epsilon_{\mathrm{dust}}$ are power-law coefficients calibrated to \Lya emitter observations in \cite{Hayes2011,Pullen2014}, which gives us a fiducial value of $f_{\mathrm{dust}}=0.65$.  However, there is considerable uncertainty in this parameter and so we will consider a range of values. We assume $f_{\mathrm{esc}}$ varies as 
\begin{equation}
    f_{\mathrm{esc}}=f_{\mathrm{esc,10}}\left(\frac{m_{\mathrm{h}}}{10^{10}\text{M}_{\odot}}\right)^{\alpha_
\mathrm{esc}},
\end{equation}
where $f_{\mathrm{esc,10}}$ is the escape fraction for a $10^{10}\text{M}_{\odot}$ halo, $m_\mathrm{h}$ is the halo mass, and $\alpha_\mathrm{esc}$ is the power-law index for the escape fraction \citep{Park2019} as used in 21cmFAST. The fiducial values we use for $f_{\mathrm{esc,10}}$ and $\alpha_\mathrm{esc}$ are 0.1 and -0.5, respectively~\cite{Yajima2011,Ferrara2013,Xu2016,Park2019}. We calculate the SFR in a manner consistent with 21cmFAST
\begin{equation}
    \SFR=\frac{M_*}{t_*H(z)^{-1}},
\end{equation}
where $t_*$ is a free parameter between zero and one, which we assume $t_*=0.5$, $H(z)$ is the Hubble parameter, and the stellar mass of the galaxy is 
\begin{equation}
    M_*=f_*\frac{\Omega_\mathrm{b}}{\Omega_\mathrm{m}}m_\mathrm{h},
\end{equation}
with $f_*$ as the star formation efficiency. To compute this efficiency we use 
\begin{equation}
    f_*=f_{*,10}\left(\frac{m_\mathrm{h}}{10^{10}\text{M}_{\odot}}\right)^{\alpha_*},
\end{equation}
where $f_{*,10}$ is the star formation efficiency for a fiducial $10^{10}$~M$_\odot$ halo and $\alpha_*$ is the coefficient to the stellar baryon power law \citep{Park2019} as used in 21cmFAST. The fiducial values we use are $f_{*,10}=0.0575$ (we include three significant figures here because the neutral fraction is sensitive to this value) and $\alpha_*=0.5$ (see~\cite{Behroozi2015,Mirocha2017,Park2019}), but we vary $f_{*,10}$ to set our ionizing efficiency. Our choices of $\alpha_{*}$ and $\alpha_\mathrm{esc}$ result in an ionizing efficiency that is independent of halo mass. 

To simplify the \Lya radiative transfer calculation, we apply a duty cycle to the galaxies so that only $10\%$ of the galaxies are emitting \Lya at a given time, but they are 10 times brighter. We find this approximation has minimal effect on our results, as it increases of power on the smallest scales by a factor of 2. We model the emergent profile of the \Lya line from each galaxy to be a Gaussian with $1-\sigma$ width of 100~km~s$^{-1}$. 

We also assume a velocity offset of 100~km~s$^{-1}$ to account for the effects of scattering in the interstellar medium (ISM), which is not resolved in our simulations. We include this because large amounts of scattering in the ISM results in an emergent spectrum sufficiently redshifted to allow photons to readily scatter to the observer and radiative transfer could be ignored. Alternatively, radiation could escape the galaxy by blueshifting instead, but redshifting has the potential to reduce the need for radiative transfer. We therefore focus on this portion of the spectrum and leave investigation of the exact emergent galaxy spectrum to future work. \cite{Blaizot2023,Smith2019,Smith2022} show the velocity offset to be between 60~km~s$^{-1}$ and 250~km~s$^{-1}$. We find that up to values of 500~km~s$^{-1}$ this velocity offset has little impact on our results, as discussed in Section~\ref{voff}. 

\subsection{Edges of Ionized Bubbles}\label{Edges}
The edges of ionized bubbles produce \Lya photons from collisional cooling of gas at the boundary of the ionized bubbles. Ionizing photons photoheat the gas which then cools along the edges of the ionizing bubbles as this region contains substantial ions and neutrals needed for efficient collisional cooling. This cooling primarily occurs via the emission of \Lya photons \citep{DAloisio2019,Pullen2014,santos2002,Wilson2024}. The \Lya luminosity from these edges is approximately proportional to the number of ionizing photons produced by each halo.

Because of the proportionality between the number of ionizing photons produced by a halo and the luminosity of the edges, we find the \Lya luminosity for the edges by calculating the number of ionizing photons produced that escapes into the IGM. The number of ionizing photons is $\dot{N}_{\mathrm{ion,\gamma}}=N_{\gamma} f_{\mathrm{esc}}\SFR/m_\mathrm{p},$ where $N_{\gamma}$=4000 is the number of ionizing photons produced per stellar baryon, and $m_\mathrm{p}$ is the mass of a proton. We then find the total \Lya luminosity of the edges to be the number of ionizing photons from the galaxy multiplied by the fraction of \Lya photons produced per ionization, $f_{\mathrm{\MathLya,e}}$. We take into account the ionizing photons required to maintain ionization in HII regions with the factor $f_{\mathrm{rec}}$ given that recombinations continually occur throughout these ionized bubbles. This gives us 
\begin{equation} \label{Ledge}
L_{\mathrm{edges}}=N_{\gamma}f_{\mathrm{\MathLya,e}}E_{\mathrm{\MathLya}}f_{\mathrm{esc}}(1-f_{\mathrm{rec}})\frac{\SFR}{m_\mathrm{p}},
\end{equation}
when summed over the SFRs of all halos. We will describe below how this is split into individual ionized bubbles. To get the number of \Lya photons produced by each ionization we assume 
\begin{equation}\label{f_esc_e}
    f_{\mathrm{\MathLya,e}}=\frac{(\epsilon-3 T_{\mathrm{gas}}k_\mathrm{b})}{10.2~\text{eV}},
\end{equation}
where $T_{\mathrm{gas}}$ is the temperature the IGM gas cools to, which we assume to be 10,000~K (on the lower end of models which suggest $10^4-2.5\times10^4$K \citep{DAloisio2019}), and $\epsilon$ is the amount of heating per ionization. In order to calculate the heating per ionization we  assume a specific flux $F_\nu\approx F_0\nu^{-\alpha}$, where $\nu$ is the frequency of the light. $F_0$ and $\alpha$ are power law coefficients for galaxy flux and we assume $\alpha=2$ for a stellar spectrum. We can then find the heating per ionization 
\begin{equation}
    \epsilon=\frac{\int^{\infty}_{\nu_0} d\nu \frac{F_\nu}{h\nu} h(\nu-\nu_{0})}{\int^{\infty}_{\nu_0} d\nu \frac{F_\nu}{h\nu}},
\end{equation}
where $\nu_{0}$ is the frequency of a photon at the threshold for ionization \cite{DAloisio2019,Wilson2024}. This expression can be simplified to $\epsilon=13.6~\text{eV}/(\alpha-1)$, as shown in \cite{DAloisio2019}. This gives us a value of $f_{\mathrm{\MathLya,e}}\approx1.08$ for our assumptions, which is higher than the values between 0.2 and 0.8 found in \cite{Wilson2024}, because of the lower value of $T_{\mathrm{gas}}$ we have assumed. We chose this lower value to bound the effect of ionization fraction emissions, which we find is likely to be subdominant even with our aggressive assumptions. We assume the IGM is sufficiently optically thick to use the case B recombination coefficient, $\alpha_\mathrm{B}$, although the difference between this assumption and case A can be absorbed into the uncertain clumping factor. The fraction due to recombinations can be approximated as 
\begin{equation}                
    f_{\mathrm{rec}}\approx C\alpha_\mathrm{B}\overline{n_\mathrm{H}}^2(1-x_{\mathrm{HI}})V_{\mathrm{box}}/\dot{N}_{\mathrm{ion,tot}},
\end{equation}
where $\dot{N}_{\mathrm{ion,tot}}$ is the total number of ionizing photons per second($\dot{N}_{\mathrm{ion,\gamma}}$) for all halos in the box, $\overline{n_\mathrm{H}}$ is the mean hydrogen number density of the universe, $C$ is the clumping factor which we assume to be 5 \citep{Iliev2007,Mcquinn2007,DAloisio2020}, $x_{\mathrm{HI}}$ is the fraction of neutral hydrogen in a cell, and $V_{\mathrm{box}}$ is the volume of our simulation box in physical units. We assume within a cell there are portions of the cell which are fully ionized and portions of the cell which are fully neutral. This assumption gives us this fraction due to recombinations. (If we instead assume the cell was uniformly partially ionized we would have an additional factor of $(1-x_\mathrm{HI})$) We also assume the \Lya line shape from bubble wall emissions to be a Gaussian with thermal broadening, giving us a $1-\sigma$ width of 9~kms$^{-1}$ corresponding to $T_\textrm{gas}=10^4$~K. This thermal broadening is calculated as $\sqrt{k_\mathrm{b}10^4~\text(K)/m_\mathrm{p}}$, where $k_\mathrm{b}$ is the Boltzmann constant.

In order to find the edges of ionized bubbles and the corresponding ionizing radiation strength, we perform a Monte Carlo ray tracing calculation. We achieve this by assigning each halo a number of rays proportional to its SFR to be emitted isotropically from each of the halos in our 21cmFAST outputs. Once the rays reach a neutral portion of the box, $x_\mathrm{HI}=1.0$, we assume this to be the edge of an ionized bubble. Each ray divides the luminosity of Eq.~\ref{Ledge} into discrete portions dependent on the halo's SFR. We use $10^7$ total rays to ensure our results are converged in our power spectrum out to a scale of $k=1$~Mpc$^{-1}$.

\subsection{Recombinations in the Intergalactic Medium}\label{Recombinations}
We use the ionization fraction and density from 21cmFAST to calculate the Ly$\alpha$ flux from recombinations. The number density of hydrogen in each cell is given by $n_\mathrm{H}=\overline{n_\mathrm{H}}(1+\delta)$ with $\delta$ being the overdensity of the cell given by the 21cmFAST simulations. We can then calculate the flux of a cell in our box due to Ly$\alpha$ from recombinations in the ionized bubbles in the IGM. We take the \Lya luminosity from recombinations to be 
\begin{equation}
L_{\mathrm{rec}}=V_{\mathrm{cell}}C\alpha_\mathrm{B} \overline{n_\mathrm{H}}^2 (1+\delta)^2(1-x_{\mathrm{HI}})f_{\mathrm{Ly\alpha,r}}E_{\mathrm{Ly\alpha}},
\end{equation}
where $V_{\mathrm{cell}}$ is the physical volume of a cell, $x_{\mathrm{HI}}$ is the hydrogen neutral fraction, $E_{\mathrm{\MathLya}}$ is the energy of a \Lya photon, and $f_{\mathrm{Ly\alpha,r}}$ is the fraction of the recombinations that will produce Ly$\alpha$ photons, which we asumme to be $0.66$~\citep{Osterbrock2006Book}. We assume the same Gaussian spectrum as we did for the edges of ionized bubbles in Section~\ref{Edges}.  

\subsection{Ultraviolet Continuum Emission from Galaxies}\label{Continuum}
An additional source of \Lya photons we consider are those produced by reprocessing of galaxy continuum emission in the IGM. As continuum emission photons leave a galaxy, some are redshifted into a Lyman series line and are absorbed in the IGM. This leads to cascading photons which have a certain probability to become Ly$\alpha$ photons depending on the Lyman series line where the absorption occurred. This physics is similar to Lyman-Werner feedback as it applies to the first stars \citep{Haiman1997,Haiman2000,Machacek2001,Wise2007,OShea2015,Ahn2009,Visbal2014}. The calculation we perform is similar to that of \cite{Ahn2009}. When we refer to an energy level, we are referring to the energy level the photon is absorbed before cascading to a \Lya photon.

These photons are reprocessed at redshift $z_{\mathrm{rep}}$, which is given by
\begin{equation}
    \frac{1+z_{\mathrm{rep}}}{1+z_{\mathrm{emit}}}=\frac{\nu_{\mathrm{rep}}}{\nu_{\mathrm{emit}}},
\end{equation}
where $z_{\mathrm{emit}}$ is the redshift of the galaxy, $\nu_{\mathrm{emit}}$ is the frequency at which the continuum photon was emitted, and $\nu_{\mathrm{rep}}$ is the Lyman series line in which the photon is reprocessed. This redshift can then be used to calculate the separation between the halo from which the photons were emitted and where they will be reabsorbed and cascade. We calculate this separation as
\begin{equation}
    r=\int^{z_{\mathrm{emit}}}_{z_{\mathrm{rep}}}c\frac{dz}{H(z)},
\end{equation}
where $c$ is the speed of light. At $z>6$, we can assume the matter-dominated case so the separation becomes
\begin{equation}
    \label{r_z_relationship} r=2cH_0^{-1}\Omega^{-1/2}_\mathrm{m}[(1+z_{\mathrm{rep}})^{-1/2}-(1+z_{\mathrm{emit}})^{-1/2}],
\end{equation}
%$
where $H_0$ is the Hubble constant. We assume the source galaxies have specific ultraviolet luminosity 
\begin{equation}
    L_\nu=a\frac{\dot{M}_*}{\text{M}_\odot\text{yr}^{-1}}\left(\frac{\nu}{\nu_{\mathrm{\MathLya}}}\right)^{-\alpha_{\mathrm{UV}}},
\end{equation}
which we normalize so we have 9690 ultraviolet photons per stellar baryon, based on the work by \cite{Barkana2005} and $\nu_{\mathrm{\MathLya}}$ is the frequency of a \Lya photon. This gives us $a=9.4637\times 10^{27}\text{erg}/\text{s}/\text{Hz}$ when we go over a wavelength range from \Lya to the Lyman limit and $\alpha_{\mathrm{UV}}=0.86$ as in \cite{Pullen2014} and is comparable to the values found in \cite{Dunlop2012,cullen2023}. Varying $\alpha_{\mathrm{UV}}$ to match the average values in these works has little impact on our final results. We can then find the flux of a galaxy at the location where continuum photons are being reprocessed to be
\begin{equation}
    \label{galaxy_flux_cont}F_{\nu}=\frac{L_{\nu}(\nu=\nu_{\mathrm{emit}})}{4\pi D_\mathrm{L}^2}\left(\frac{1+z_{\mathrm{emit}}}{1+z_{\mathrm{rep}}}\right)\exp{[-\tau_{\nu}]},
\end{equation}
where $\tau_\nu$ is the optical depth of that cell. We assume the optical depth is zero everywhere except for at the Lyman series lines, which we assume to be infinite. We can make this assumption because the optical depth of a Lyman series line is very high at line center at relevant redshifts. This assumption is less accurate for $n>3$, but these higher energy levels are a much smaller portion of the spectrum~\citep{Yang2020}. The luminosity distance $D_\mathrm{L}$ is given by
\begin{equation}
    D_\mathrm{L}\equiv\left(\frac{r}{1+z_{\mathrm{rep}}}\right)\left(\frac{1+z_{\mathrm{emit}}}{1+z_{\mathrm{rep}}}\right).
\end{equation}
Using Eq.~\ref{galaxy_flux_cont}, we calculate the Ly$\alpha$ luminosity $L_{\mathrm{Ly\alpha,n}}$ from continuum emission in each cell that are absorbed at energy level $n$ before cascading to be
\begin{equation}
\begin{split}
    L_{\mathrm{Ly\alpha,n}}=&\frac{E_{\mathrm{Ly\alpha}}}{E_\mathrm{n}}F_\nu f_{\mathrm{Ly\alpha,n}}A_{\mathrm{pixel}}\Delta\nu\\
     =&\frac{\nu_{\mathrm{Ly\alpha}}}{\nu_{\mathrm{rep}}}\frac{L_\nu(\nu=\nu_{\mathrm{emit}})}{4\pi}\left(\frac{1+z_{\mathrm{rep}}}{r}\right)^2\left(\frac{1+z_{\mathrm{rep}}}{1+z_{\mathrm{emit}}}\right) f_{\mathrm{Ly\alpha,n}}A_{\mathrm{pixel}}\Delta\nu,\\
     \label{continuumAnalytic}
\end{split}
\end{equation}
where $f_{\mathrm{Ly\alpha,n}}$ is the probability the photon will cascade into Ly$\alpha$ for each energy level, given by \cite{Hirata2006}, $A_{\mathrm{pixel}}$ is the area of the pixel, $E_\mathrm{n}$ is the energy corresponding to energy level $n$, and $\Delta\nu$ is the frequency a photon would redshift across a cell a distance $r$ away from the source if it was not absorbed at a Lyman-series line. The total luminosity from the continuum in a cell, $L_{\mathrm{cont}}$, is the sum of Eq.~\ref{continuumAnalytic} from $n=2$ to $n=30$. We only consider energy levels up to $n=30$ as this yields convergent results in our luminosity. Note that Eq.~\ref{continuumAnalytic} assumes we are working far from the source of the continuum.

\begin{figure}
    \centering
    \includegraphics[width=.6\textwidth]{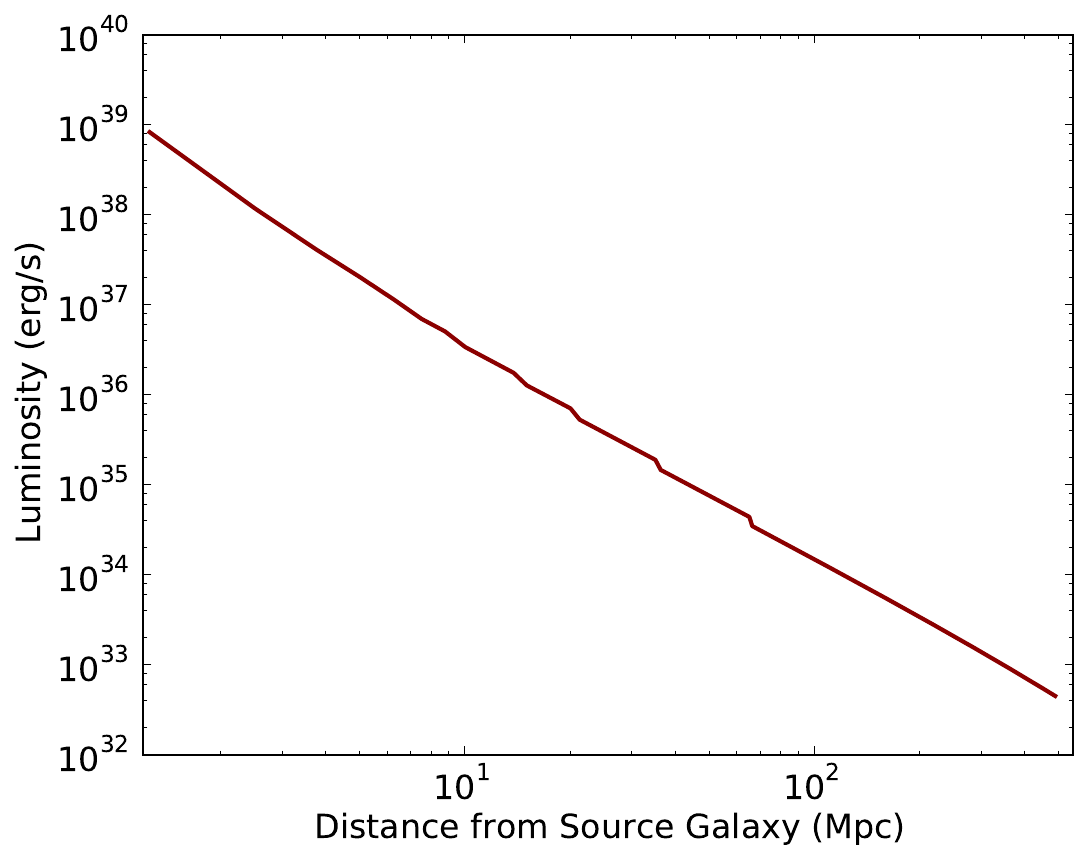}
    \caption{The \Lya luminosity in one of our simulation cells from the continuum as a function of distance from the source galaxy. The steps in this function are the result of the different distances associated with each energy level.}
    \label{fig:lum_r_relation}
\end{figure}

When values of $r$ are small, we must be more precise in our solution for finding the reprocessing to avoid unphysical solutions. To do this we perform a volume integral of Eq.~\ref{continuumAnalytic}. We find the equation for the luminosity $L_{\mathrm{\MathLya,n}}$ of a cell to be
\begin{equation}\label{ContinuumFull}
    \begin{split}
     L_{\mathrm{\MathLya,n}}(r)=&\int_V \frac{E_{\mathrm{\MathLya}}}{E_\mathrm{n}}F_\nu f_{\mathrm{Ly\alpha,n}}\frac{d\nu_{\mathrm{emit}}}{dr}\,r^2 dr \sin\theta\,d\theta\,d\phi,
     \\=&\frac{E_{\mathrm{Ly\alpha}}f_{\mathrm{Ly\alpha,n}}}{4\pi h}\frac{1}{(1+z_{\mathrm{emit}})}\int_V \frac{L_\nu(\nu_{\mathrm{emit}})(1+z_{\mathrm{rep}})}{\nu_{\mathrm{emit}}}\frac{d\nu_{\mathrm{emit}}}{dr}\,dr \sin\theta\,d\theta\,d\phi,
    \end{split}
\end{equation}
where $h$ is Planck's constant. The bounds of this integral are the edge of the cell at a distance $r$ from the originating halo. We solve this integral numerically for many distances from the originating halo and then sum over all energy levels. We also normalize this equation to be for a halo with a SFR $1~\text{M}_\odot\text{yr}^{-1}$. This gives us the total continuum luminosity of a single halo as a function of distance as shown in Figure~\ref{fig:lum_r_relation}.

To get the total continuum luminosity from all of the halos in the box we convolve the luminosity as a function of $r$ with the data cube from 21cmFAST of total SFRD of each cell ($\dot{\rho}_\mathrm{*,cell}$). This gives us $L_{\mathrm{cont,tot}}=L_{\mathrm{\MathLya}}(r)*\dot{\rho}_\mathrm{*,cell}(r)/(\text{M}_\odot\text{yr}^{-1})$, where $L_{\mathrm{\MathLya}}$ is the sum of Eq.~\ref{ContinuumFull} across all energy levels, is normalized to a SFR of $1~\text{M}_\odot\text{yr}^{-1}$. We have shown $L_{\mathrm{\MathLya}}$ in Figure~\ref{fig:lum_r_relation} and this function is spherically symmetric. We have included the convolution equation as the continuous convolution for clarity, but perform a discrete convolution in the simulation. This includes periodic boundary conditions and the SFR is calculated the same way as was done in section \ref{Halos}. 

The maximum distance that continuum photons travel from a given halo is 490~Mpc which exceed the size of our simulation box. This distance is set by how far a photon just outside of line center of $n=3$ would travel to redshift into the line center of $n=2$. In order to account for the continumm from halos outside the box we have to estimate their contribution. To do this we calculate the SFR in spherical shells with a radius greater than 100~Mpc (half the size of our simulation box). Within each spherical shell we estimate the abundance of halos of a given mass, $dn/dM$, using the Sheth-Tormen halo mass function \citep{Sheth1999}. We calculate the redshift of the shell using Eq.~\ref{r_z_relationship} solved for $z_{\mathrm{emit}}$ in order to calculate the mass function at that redshift. We then calculate the star formation rate density (SFRD), 
\begin{equation}
    \SFRD=\int \frac{dn}{dM} \SFR dM.
\end{equation}
where the SFR is calculated in the same way as before. From this we can then calculate the SFR of the shells, $\SFR_{\mathrm{shell}}=\SFRD V_{\mathrm{shell}},$ where $V_{\mathrm{shell}}$ is the volume of the spherical shell.

Using the SFR of each shell we can find the corresponding luminosity the shell deposits onto one cell, using the relationship in Eq.~\ref{ContinuumFull}. We  numerically integrate these shells to get the total luminosity deposited in a cell to be 
\begin{equation}
    L_{\mathrm{hor}}=\int^{490~\text{Mpc}}_{100~\text{Mpc}}\SFRD  L(r_\mathrm{shell}) 4\pi r^2dr,
\end{equation}
where $L(r_\mathrm{shell})$ is normalized for SFR of $1\text{M}_{\odot}\text{yr}^{-1}$. This equation only includes $n=2$ since this is the only Lyman series line that requires a distance of 100~Mpc or greater. We then apply this estimate to each cell in our simulation box so that we account for how the continuum emission from halos outside our simulation box affects the intensity map. We find this to be $\approx40\%$ of the total continuum signal. 

For our calculations, we assumed a Gaussian spectrum with thermal broadening for all components of the continuum. This spectrum used is the same as in Sections~\ref{Edges} and \ref{Recombinations}.

\subsection{\Lya Radiative Transfer}\label{RadTrans}

The radiative transfer code we use is a Monte Carlo radiative transfer code similar to \cite{Faucher2010} and was used in \cite{Visbal2018} (see also \cite{Zheng2002,Cantalupo2005,Dijkstra2006,Laursen2007}). We divide the luminosity of the sources into discrete photon packets. These photons will scatter and take a random walk in both frequency and position space until they escape the simulation box. At each scattering we determine the probability of a photon scattering to the observer and use this as the contribution to our intensity maps.

For each run we use three million photons. We check this convergence using the power spectra of the resulting intensity maps and find this many photons gives us convergence to a scale of $k=1$~Mpc$^{-1}$. 

\section{Results} \label{sec:results}

Here we present our simulated intensity maps as well as their associated power spectra. 
In Section~\ref{SourcesonIM} we show each source of \Lya photons. In Section~\ref{AstroParameters} we show how different parameters effect the power spectra, including the neutral fraction, the fraction of \Lya absorbed by dust, and ionizing photon escape fraction. Our tested neutral fractions include $x_\mathrm{HI}=0.21, 0.5, 0.75$. Our dust models include $f_\mathrm{dust}=$ 0.99, 0.65, 0.3, and 0 for all three neutral fractions tested. Our models for escape fraction include $f_\mathrm{esc,10}=$ 0.1, 0.3, and 0.8 with $x_\mathrm{HI}=0.5$. In Section~\ref{voff}, we show how various values for our velocity offset, used to approximate radiative transfer in the ISM, affects our simulation. In Section~\ref{cross_corrrelation} we show a cross correlation between these intensity maps and a hypothetical galaxy survey alongside sensitivity curves for SPHEREx and a CDIM-like instrument. Our fiducial model has the parameters $z=7.0$, $f_\mathrm{esc,10}=0.1$, $x_\mathrm{HI}=0.5$, and $f_\mathrm{dust}=0.65$.

\subsection{Sources of \Lya Photons} \label{SourcesonIM}

\begin{figure*}
    \centering
    \includegraphics[width=1\linewidth]{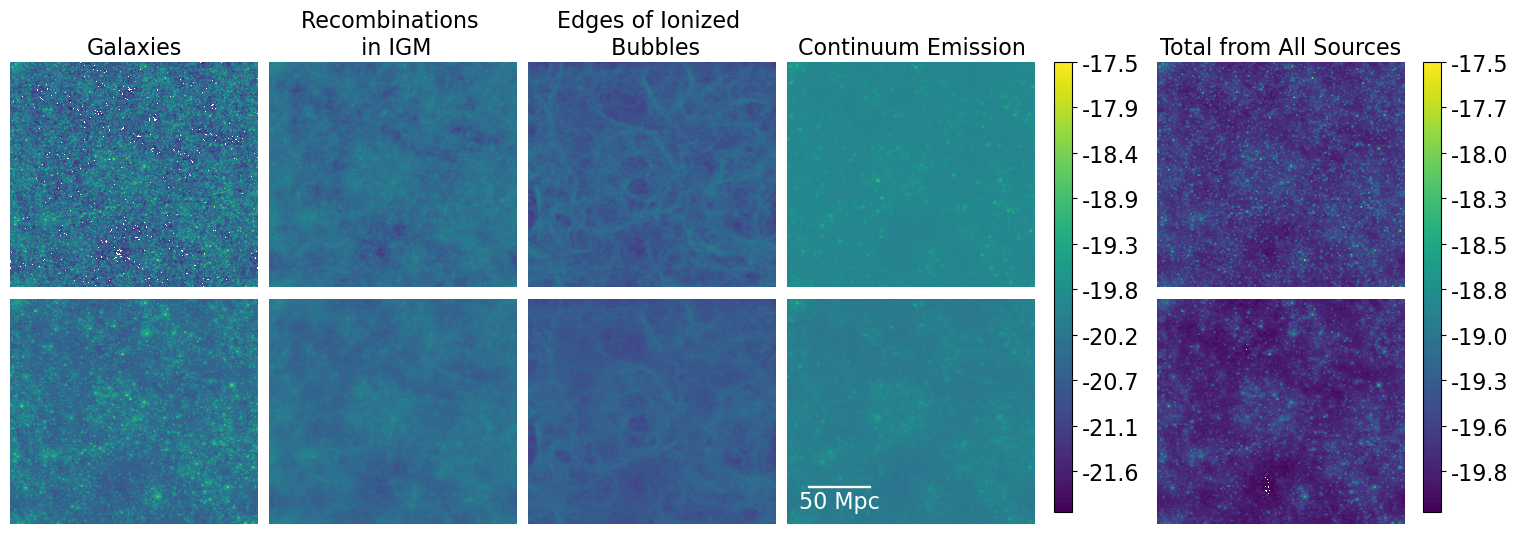}
    \caption{Intensity maps of each source with (bottom) and without (top) \Lya radiative transfer at $z=7.0$, a neutral fraction of $0.50$, escape fraction $f_\mathrm{esc,10}=0.1$, dust absorption $f_\mathrm{dust}=0.65$. The right column shows the total from accounting for all four sources of Ly$\alpha$. The color bars are in units of $\log(\text{nWm}^{-2}\text{sr}^{-1})$ and the maps are 200~Mpc on each side.  
    }
    \label{fig:fluxmap_fiducial}
\end{figure*}

In Figure~\ref{fig:fluxmap_fiducial}, we show the intensity maps for each source from our fiducial model before and after applying the \Lya radiative transfer. We see when we include \Lya radiative transfer the signal is smeared out. The maps are visually dominated by the continuum emission and \Lya photons from galaxies regardless of the inclusion of radiative transfer.

\begin{figure}
    \centering
    \includegraphics[width=1\textwidth]{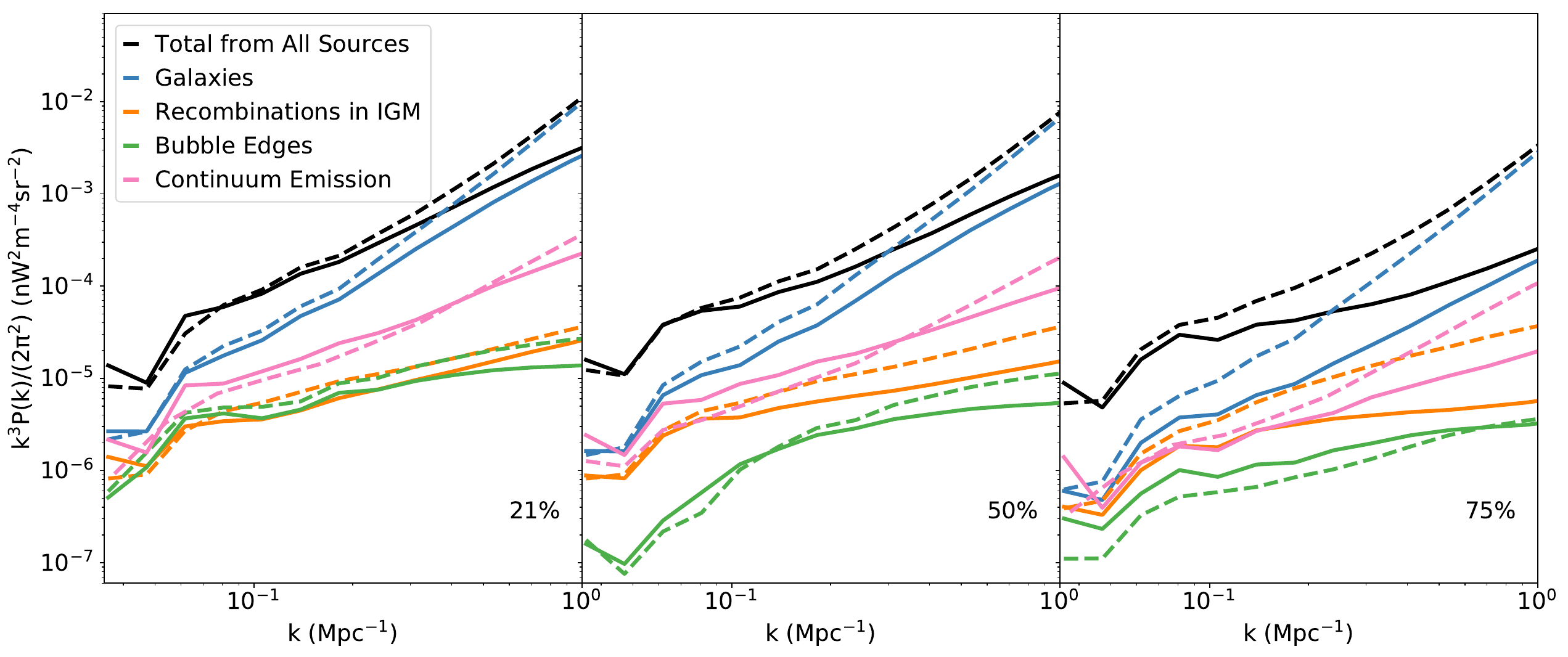}
    \caption{Power spectra for intensity maps with three different neutral fractions with $z=7.0$, $f_\mathrm{esc,10}=0.1$, and $f_\mathrm{dust}=0.65$. From left to right, the neutral fraction is 21\%, 50\%, and 75\%. Solid and dashed lines respectively depict simulations with and without radiative transfer. The black is the total from all sources of Ly$\alpha$, the blue is the \Lya emissions from galaxies, the orange is the \Lya from recombinations in the IGM, the green is the \Lya from cooling at the edges of ionized bubbles, and the pink is continuum emission cascading to \Lya photons. The radiative transfer has a large impact on the power spectrum, particularly on small scales, and has a larger impact the higher the neutral fraction of the IGM. }
    \label{fig:power_diff_nf_3_panel}
\end{figure}

In Figure~\ref{fig:power_diff_nf_3_panel}, we show the power spectra of the different \Lya sources for three different neutral fractions: 21\%, 50\% (fiducial), and 75\%. The neutral fraction is dependent on the SFR and the escape fraction of ionizing radiation. In order to obtain these neutral fractions we changed the SFR of the galaxies by varying the star formation efficiency parameter $f_{*,10}=$0.07, 0.0575, and 0.0375 for 21\%, 50\%, and 75\%, respectively, while holding the escape fraction of ionizing radiation constant. 
The power of the galaxies and continuum decreases with decreasing SFR. The power of the recombinations and edges also decrease as the neutral fraction is increased due to the change in bubble volume and surface area, respectively.

In all panels of Figure~\ref{fig:power_diff_nf_3_panel}, the \Lya photons directly from galaxies contribute the most power. The continuum emission, however, becomes comparable, particularly on large scales, to the galaxies when radiative transfer is included. Recombinations and edges have a lower amplitude of power on all distance scales, regardless of radiative transfer. However, their relative contributions changes, varying with neutral fraction. We will discuss what parameters cause the power of the recombinations and edges to be higher than halos and continuum in following sections. It is important to note the total power spectrum is from taking the modulus squared of the sum of all contributions to the \Lya intensity, not the sum of the power spectra of the components (i.e. there are additional cross-power spectra that reflect correlations between the different \Lya sources).

\subsection{Astrophysical Parameter Effects}\label{AstroParameters}

In this section we examine how varying three different astrophysical parameters --- the neutral fraction, the dust absorption in galaxies, and the ionizing photon escape fraction --- affects our results. We also look at how this impacts the power of each source of \Lya photons. 

\subsubsection{Effect of Neutral Fraction} \label{nfEffect}

\begin{figure}
    \centering
    \includegraphics[width=0.6\textwidth]{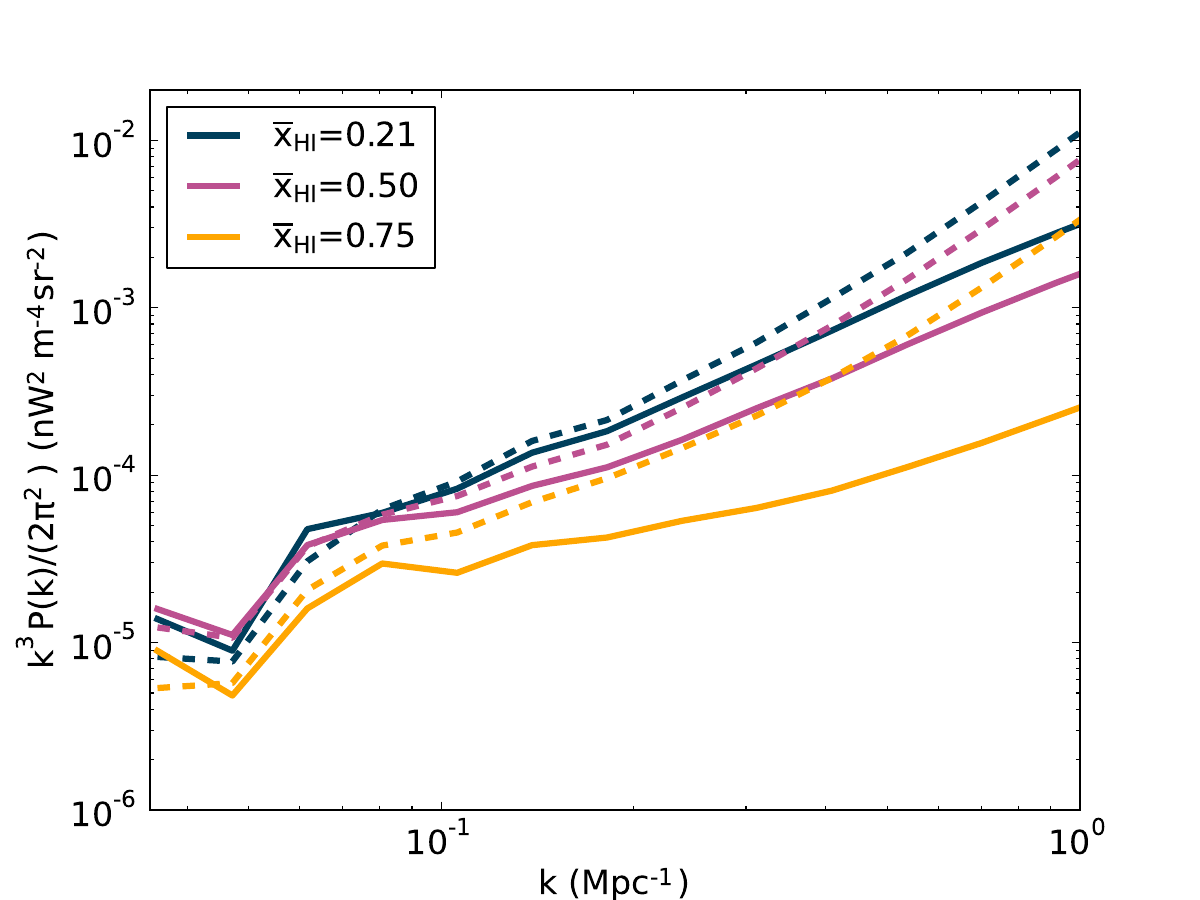}
    \caption{The impact of various neutral fractions on \Lya power spectrum. We show simulations with (solid) and without (dashed) including radiative transfer, for neutral fractions of 21\% (navy), 50\% (purple), and 75\% (orange). Radiative transfer has a larger effect on the slope at higher neutral fractions.}
    \label{fig:nf_power_compare}
\end{figure}

We show the total power spectra from all sources for our three neutral fractions in Figure~\ref{fig:nf_power_compare}. We can see that the neutral fraction has an effect on how much the slope of the power spectrum is decreased with radiative transfer. As shown in \cite{Visbal2018}, \Lya radiative transfer smears out signal up to $\approx10$~Mpc. This smearing more heavily affects the signal on smaller scales causing a downward tilt in slope for our power spectra. For 21\% neutral, $P_\mathrm{\MathLya}\propto k^{-1.4}$. For 50\% neutral, $P_\mathrm{\MathLya}\propto k^{-1.7}$. For 75\% neutral, $P_\mathrm{\MathLya}\propto k^{-2.0}$. (Before radiative transfer 21\% neutral $P_\mathrm{\MathLya}\propto k^{-1.15}$, 50\% $P_\mathrm{\MathLya}\propto k^{-1.33}$, and 75\% neutral $P_\mathrm{\MathLya}\propto k^{-1.46}$.) All of these are fit $0.1<k<1.0$~Mpc$^{-1}$ with a least-squared fit. We chose to begin our fit at $k=0.1$ to prevent the dip centered at $k\approx0.05$ (from cosmic variance) affecting our fit. We cut off at $k=1.0$ because this is where shot noise of Monte Carlo photons in our simulation is negligible. From Figure~\ref{fig:nf_power_compare}, we see that the slope is sensitive to the neutral fraction, even with all major sources of Lya considered. We also note that for the fiducial model we find an amplitude of $P_\mathrm{\MathLya}=0.99-0.035$~nW$^2$m$^{-4}$sr$^{-2}$Mpc$^3$ for $k\approx0.1-0.9$~Mpc$^{-1}$.

In Figure~\ref{fig:nf_power_compare}, we also see that on large scales the power is comparable between models with different neutral fractions. This is important because it shows how the power on such scales is relatively unaffected by changing SFR, so the slope is instead dependent on the neutral fraction. The uniform changing in power across distance scales is because the mean \Lya luminosity from galaxies and the continuum emission depends on the SFR. 

We see some degeneracies between the slope changing due to the neutral fraction and the slope changing from other astrophysical parameters. We will discuss these degeneracies further in Sections~\ref{DustEffect}~and~\ref{EscapeFraction}. 

\subsubsection{Effect of Dust Absorption} \label{DustEffect}

\begin{figure}
    \centering
    \includegraphics[width=1\textwidth]{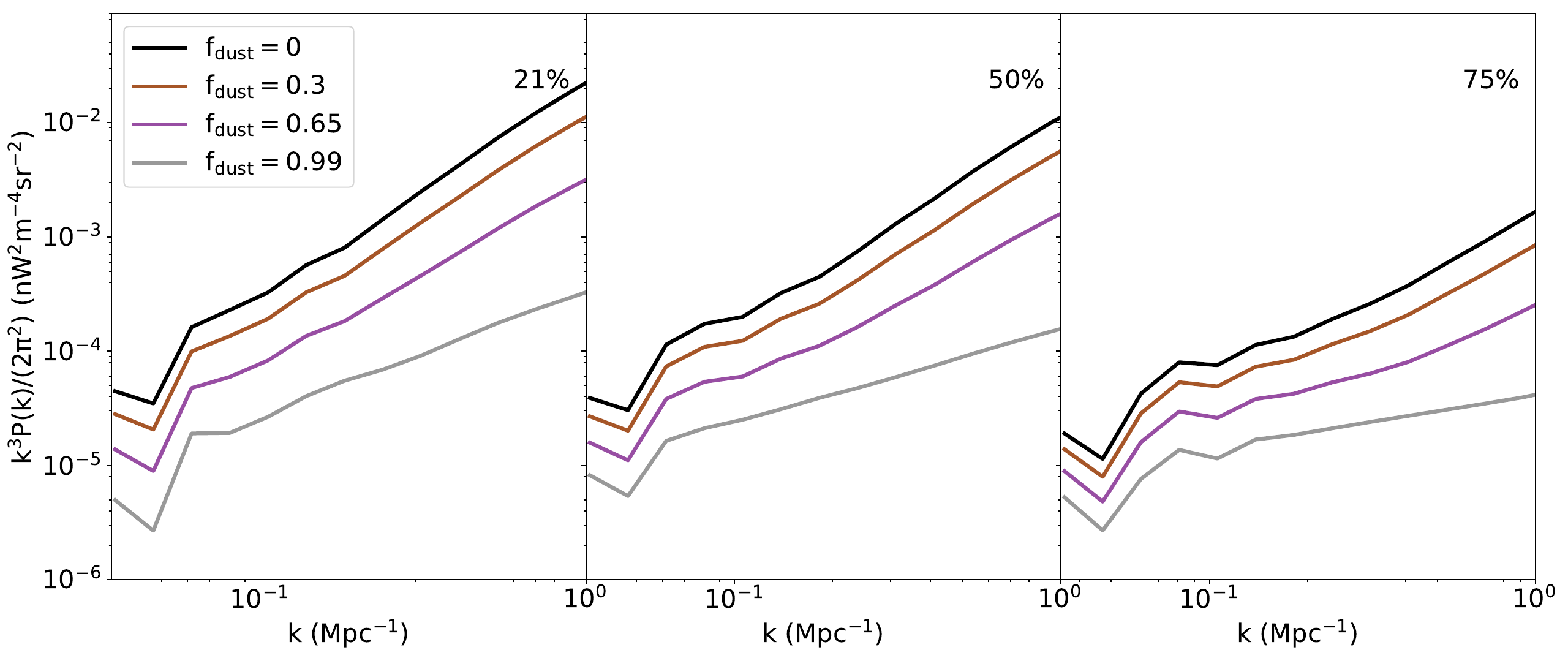}
    \caption{Power spectra for three different neutral fractions 21\%, 50\%, and 75\% (left to right) with four different dust absorption fractions 0 (black), 0.3 (brown), 0.65 (purple, fiducial model), and 0.99 (gray). While dust absorption changes the slope of the total power spectra, this effect is subdominant to the effect of different neutral fractions.}
    \label{fig:power_dust_3_panel}
\end{figure}

We investigate the effect the fraction of \Lya photons absorbed by the dust in a galaxy has on our power spectra.  As mentioned in Section~\ref{Halos}, we assume that dust only absorbs \Lya photons and not continuum photons because of scattering of \Lya in the galaxy. Figure~\ref{fig:power_dust_3_panel} shows the power spectra for the three different neutral fractions, with each panel showing different dust absorption fractions of 0, 0.3, 0.65 (fiducial model), and 0.99. The slope of the power spectra decreases with higher dust absorption because it suppresses power from galaxies while leaving the continuum emission unaffected. The continuum power spectrum is flatter than the galaxies, so larger contributions from the continuum result in a flatter power spectrum. We also see higher dust absorption results in suppression of power on all scales. 

We note, we only see a substantial change in the slope from the highest value of dust absorption. For $x_\mathrm{HI}=0.5$ and $f_{dust}=0$, $P_\mathrm{\MathLya}\propto k^{-1.3}$, for $f_\mathrm{dust}=0.3$, $P_\mathrm{\MathLya}\propto k^{-1.5}$, and for $f_\mathrm{dust}=0.99$, $P_\mathrm{\MathLya}\propto k^{-2.1}$. We expect the neutral fraction to evolve more rapidly than the dust obscuration, so taking intensity maps over a few redshifts could disentangle this degeneracy and allow us to infer the neutral fraction from the power spectrum slope. 

\subsubsection{Effect of Ionizing Photon Escape Fraction} \label{EscapeFraction}

\begin{figure}
    \centering
    \includegraphics[width=1\textwidth]{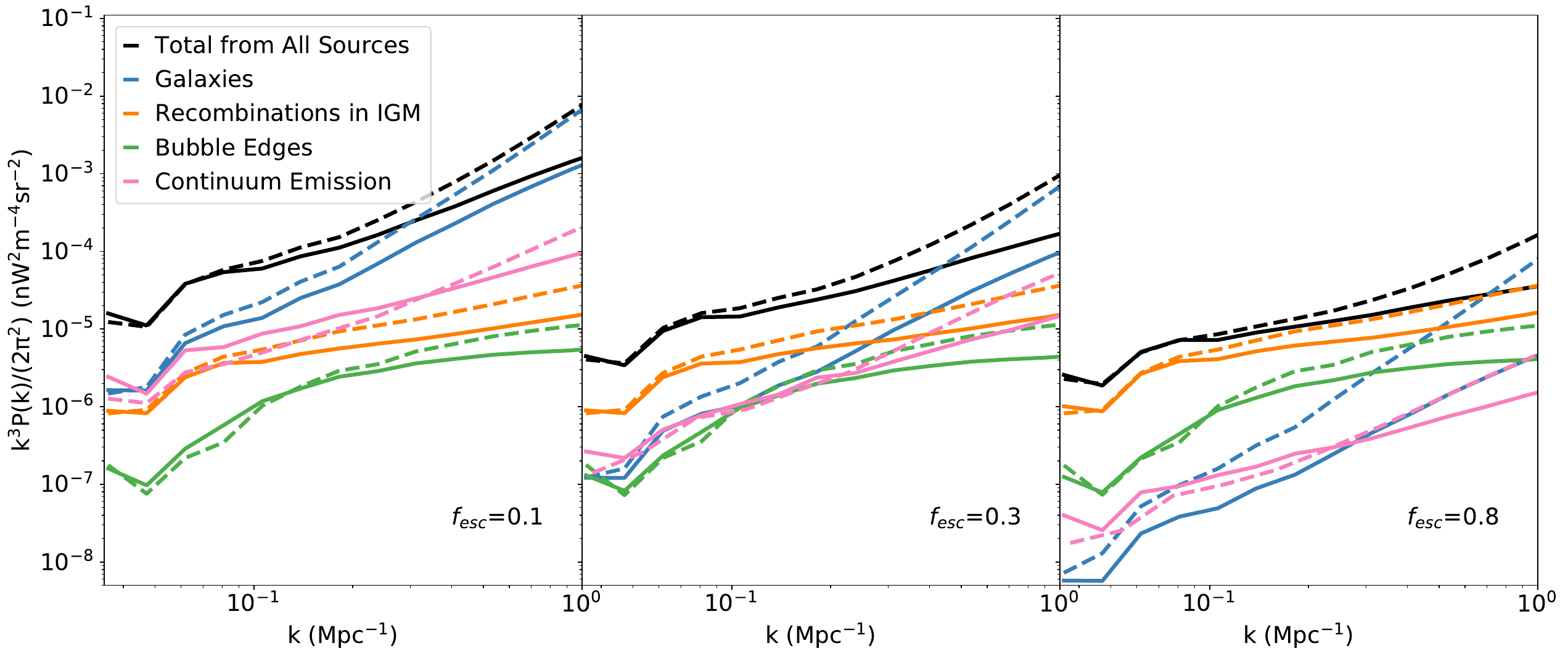}
    \caption{Power spectra for three different ionized photon escape fractions 0.1, 0.3, and 0.8 (left to right). The neutral fractions is held at $x_\mathrm{HI}=0.5$, which means we must decrease the SFR as we increase the escape fraction. Depending on the escape fraction, different sources of \Lya have higher amplitudes of power than others. For example, as we raise the escape fraction the recombinations and edges have higher amplitudes than the halos and continuum. The line style and colors on this plot are the same as in Fig.~\ref{fig:power_diff_nf_3_panel}.}
    \label{fig:power_diff_fesc_3_panel}
\end{figure}

The last parameter we vary in our models is the ionizing photon escape fraction. In Figure~\ref{fig:power_diff_fesc_3_panel} we show the results with $f_\mathrm{esc,10}=$0.1, 0.3, and 0.8 with a neutral fraction $x_\mathrm{HI}=0.5$. In order to vary the escape fraction, while hold the neutral fraction constant, we vary our star formation efficiency parameters $f_{*,10}=$0.0575, 0.0192, and 0.00719 for $f_\mathrm{esc,10}=$0.1, 0.3, and 0.8 respectively. The amplitude of the power from each \Lya source depends on the escape fraction. For our lowest escape fraction, the galaxies and continuum have higher power than the other sources because the power for the galaxies proportional to $(1-f_\mathrm{esc})$ and the power for the edges and combinations proportional to $f_\mathrm{esc}$. The highest escape fraction results in higher power from recombinations, followed by the bubble edges. 

\begin{figure}
    \centering
    \includegraphics[width=0.6\textwidth]{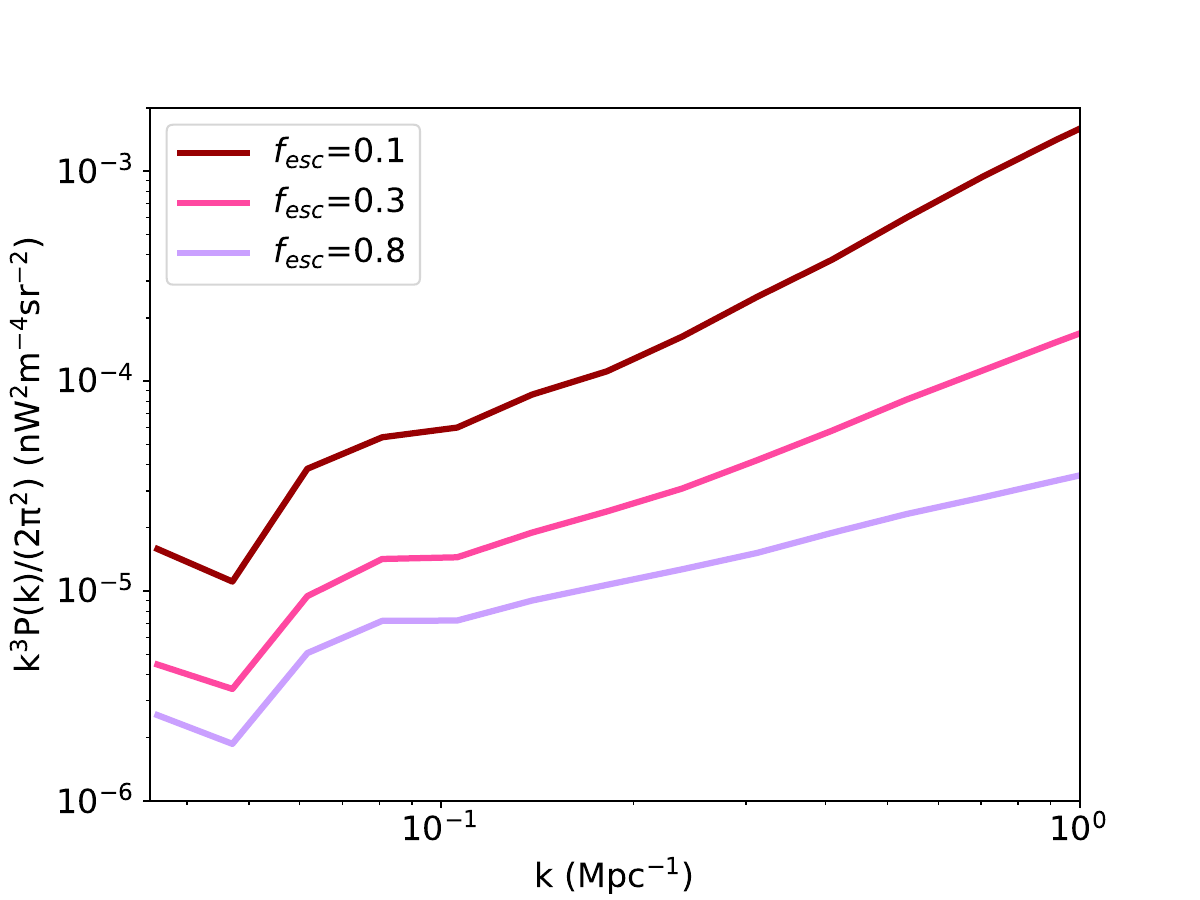}
    \caption{Total power spectra for the three different escape fractions with a neutral fraction $x_\mathrm{HI}=0.5$. We consider escape fractions of 0.1 (red), 0.3 (pink), and 0.8 (purple). Increasing the escape fraction decreases the slope of the power spectrum.}
    \label{fig:power_fesc_compare}
\end{figure}

Figure~\ref{fig:power_fesc_compare} shows the total power spectra for the three different escape fraction values and a neutral fraction of $50\%$. The slope of the power decreases with increasing escape fraction. For $f_\mathrm{esc}=0.3$, $P_\mathrm{\MathLya}\propto k^{-2.0}$ and for $f_\mathrm{esc}=0.8$, $P_\mathrm{\MathLya}\propto k^{-2.3}$. We see the slope heavily affected by different values of the escape fraction. However, we believe that the neutral fraction would evolve much more rapidly, as was the case with dust absorption, so by taking intensity maps at multiple redshifts we could disentangle the degeneracy between these parameters. We save this calculation for future work.

\subsection{Velocity Offset} \label{voff}

\begin{figure}
    \centering
    \includegraphics[width=0.6\textwidth]{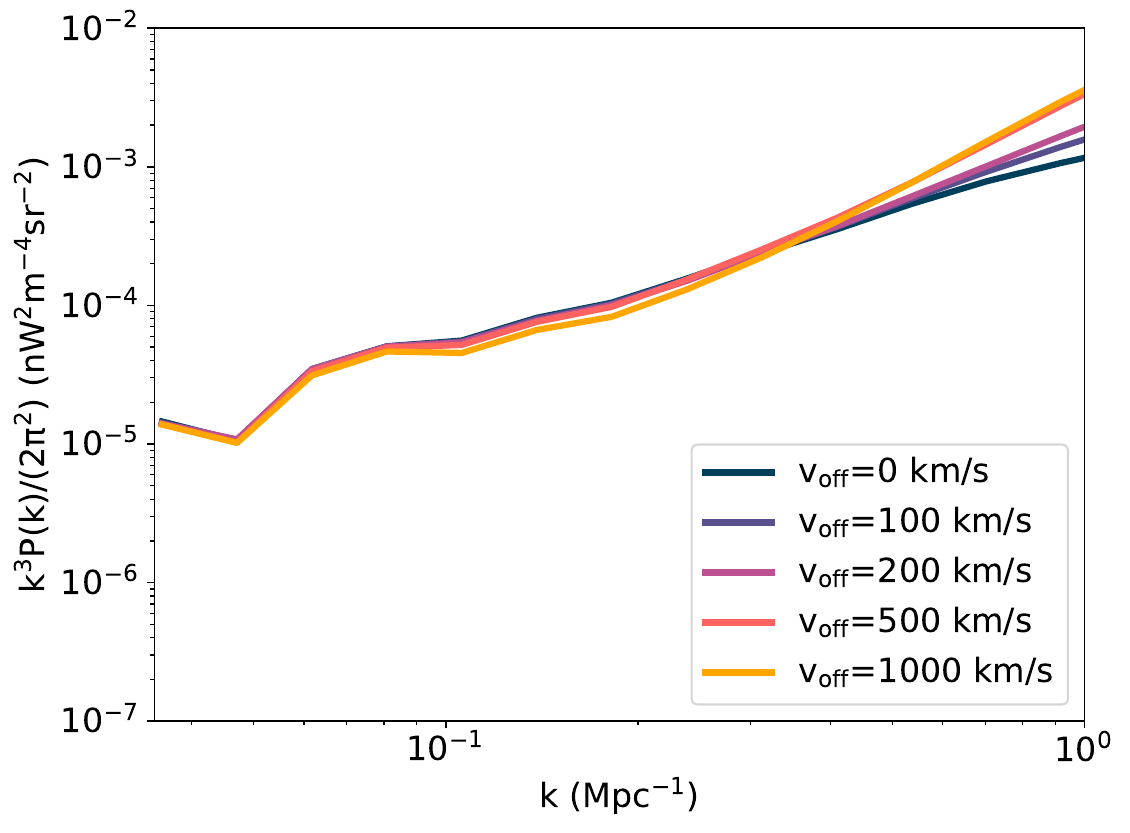}
    \caption{Power spectra for five different velocity offset values for the galaxies. The navy is 0 km/s, the purple is 100 km/s (fiducial), the pink is 200 km/s, the orange is 500 km/s, and the yellow is 1000 km/s. From this we can see there is a minimal deviation from our fiducial result, with more significant changes requiring offsets of $>500$~km/s.}
    \label{fig:power_voff}
\end{figure}

Figure~\ref{fig:power_voff} shows how changing the velocity offset value for our galaxies alters the power spectra. We include this offset to approximate radiative transfer from the ISM which is not resolved in our simulations. We can see that there is minimal change in the power spectrum from including this velocity offset for values $<500$~kms$^{-1}$ as the photons are still close enough to the \Lya line to readily scatter. This shows radiative transfer in the ISM does not negate the need for including radiative transfer in the IGM for all values reported in \cite{Blaizot2023,Smith2019,Smith2022}. If we instead include a symmetric profile of two Gaussians at $+/-100$~km/s we will find a result comparable to that with no velocity offset.

\subsection{Cross Correlation with a Galaxy Survey} \label{cross_corrrelation}

\begin{figure}
    \centering
    \includegraphics[width=0.6\textwidth]{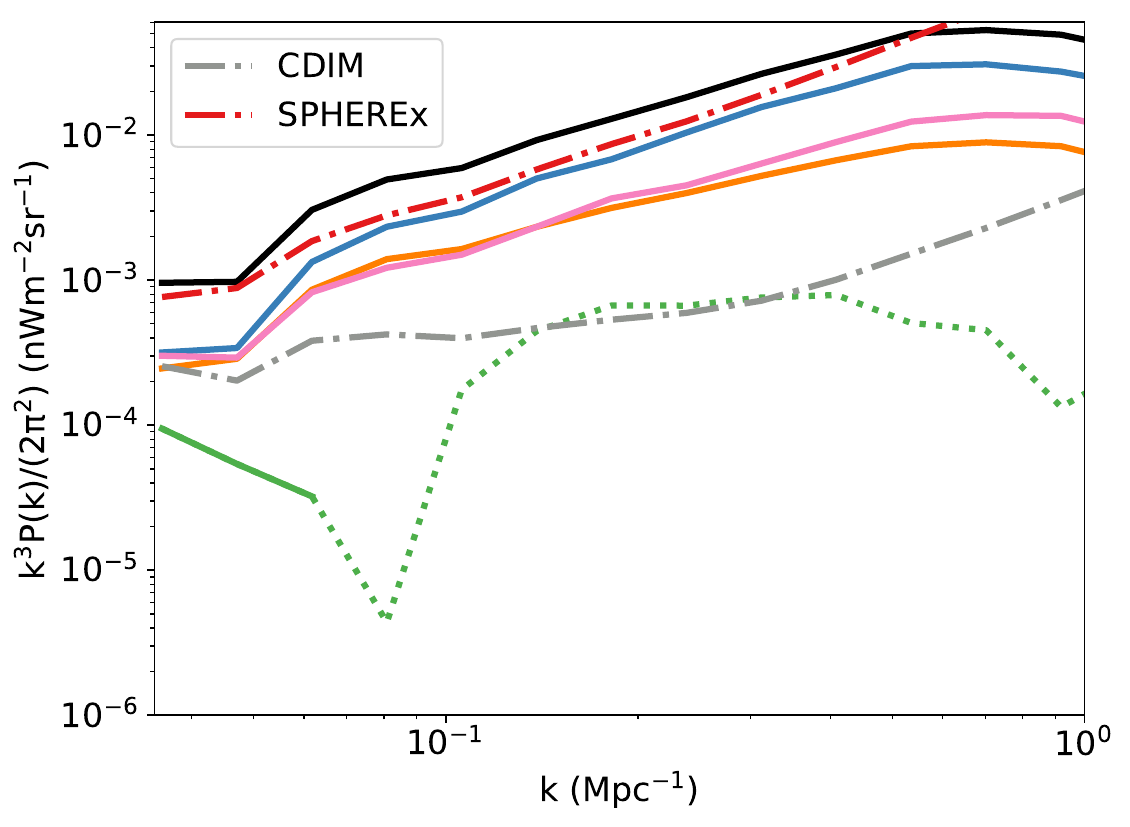}
    \caption{Cross power spectrum between our simulated intensity maps and a hypothetical galaxy survey. Line colors for the total, galaxies, continuum, recombinations, and edges power spectrum are the same as in Fig.~\ref{fig:nf_power_compare}. The dotting on the edges line indicates anticorrelation. The dot-dashed lines are sensitivity curves for SPHEREx (red) and a CDIM-like telescope (gray). These show that our total power will be observable with a total S/N$\approx$4 for SPHEREx and S/N$\approx$12 for CDIM.}
    \label{fig:cross_power}
\end{figure}

One of the main challenges to \Lya intensity mapping during reionization are interloping H$\alpha$ sources. Intensity maps of \Lya targeting $z=7$ will contain foreground signal from H$\alpha$ emitting galaxies at $z\approx 0.5$. As a result, getting an auto power spectrum for \Lya will be challenging if not impossible \citep{Pullen2014}. For this reason, we investigate a cross correlation with a hypothetical galaxy survey in Figure~\ref{fig:cross_power}. For this galaxy survey, we assume all galaxies with a mass $m_\mathrm{h}>10^{11}\text{M}_{\odot}$ are observed. This gives us a volume density of $n_\mathrm{gal}=4.5\times 10^{-4}$~Mpc$^{-3}$, similar to the \textit{Subaru} Deep field narrow band survey at $z=6.7$~\citep{Ouchi2018}, but would require more accurate redshift measurements than what is currently available for this data set. For this survey, we assume a $3^{\circ}\times 3^{\circ}$ field of view on the sky and $\Delta z=0.5$ along the line of sight based on ~\cite{Higuchi2019}; however, we note that this work covered $z=6-6.574$ while our work is focused on $z=7$. We use our 21cmFAST halo simulations in order to create synthetic galaxy redshift surveys with these specifications, and leave the investigation of more realistic surveys to future work. Our resulting fiducial model total power spans a dynamic range of $P_\mathrm{\MathLya,gal}=98.5-1.25$~nWm$^{-2}$sr$^{-1}$Mpc$^{3}$ for $k\approx0.1-0.9$~Mpc$^{-1}$.

In Figure~\ref{fig:cross_power}, we also include sensitivity curves for SPHEREx and a CDIM-like telescope. For these, we find the error on the cross power to be 
\begin{equation}
    \delta P_\mathrm{\MathLya,gal}(k)^2=\frac{\frac{1}{2}(P_\mathrm{\MathLya,gal}^2+P_\mathrm{\MathLya}P_\mathrm{gal})}{N_\mathrm{k}},
\end{equation}
where $P_\mathrm{\MathLya,gal}$ is the cross power between our simulation and a hypothetical galaxy survey, $P_\mathrm{gal}$ is the auto power from this galaxy survey, $N_\mathrm{k}$ is the number of modes in a k-bin, and $P_\mathrm{\MathLya}$ is the combination auto power from the simulation, noise power from SPHEREx or CDIM, and the power from the interloper H$\alpha$,  $P_\mathrm{\MathLya}=P_\mathrm{auto,sim}+P_\mathrm{N}+P_\mathrm{H\alpha}$. 

For SPHEREx, we assume the noise power to be $P_\mathrm{N}=\sigma_\mathrm{N}^2 \Omega_\mathrm{pix}$, where the instrument noise is assumed to be Gaussian with variance $\sigma_\mathrm{N}^2$ and $\Omega_\mathrm{pix}$ is the size of a pixel for observations. We use
$\sigma_\mathrm{N}=3$~nW~m$^{-2}$~sr$^{-1}$ from \cite{Cheng2022}. For a CDIM-like instrument, we assume 1.5~m telescope diameter and a spectral resolution of R=300 \citep{Cooray2019} and calculate the noise power in the same way as \cite{Comaschi2016}. We also assume an integration time of $10^5$~s, an observation efficiency from instrument losses of $\epsilon=0.5$, and a zodiacal light background intensity of $\nu I_{\nu}=500$~nW~m$^{-2}$sr$^
{-1}$ \citep{Pullen2014}. 

We find this cross powerp is observable with SPHEREx up to $k\approx 0.8$, with a total signal-to-noise S/N$\approx$4 from $k=0.035-1~$Mpc$^{-1}$ with bins $\Delta k=k/5$. The total signal-to-noise is $\text{S/N}=\left(\sum_\mathrm{k} (P_\mathrm{\MathLya,gal}/\delta P_\mathrm{\MathLya,gal}\right)^{1/2}$. We only consider $k<1~$Mpc$^{-1}$ because this is when the shot noise of the Monte Carlo photons in our simulation are negligible. For CDIM, we find S/N$\approx$12 across all modes from $k=0.035-1~$Mpc$^{-1}$ with bins $\Delta k=k/5$.

\begin{figure}
    \centering
    \includegraphics[width=1.0\textwidth]{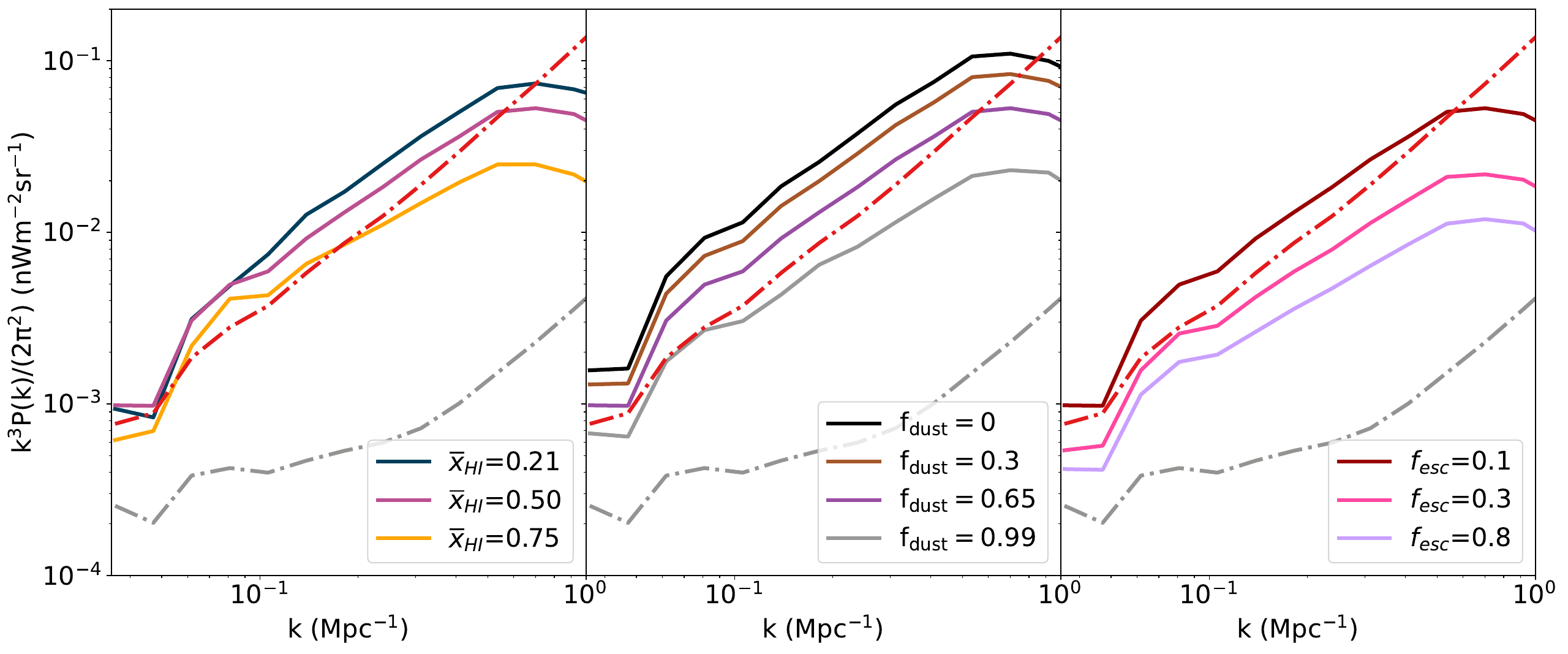}
    \caption{Dependence of cross power for each of the astrophysical parameters we varied with our hypothetical galaxy survey. From left to right the panels show the three neutral fractions, the four dust absorption fractions, and the ionizing escape fraction. The colors of this plot are the same as in other figures comparing these astrophysical parameter changes and the CDIM (gray) and SPHEREx (red) sensitivity curves the dot-dashed lines}
    \label{fig:power_all_cross}
\end{figure}

Figure~\ref{fig:power_all_cross} shows the cross power for this hypothetical galaxy survey with varying astrophysical parameters, including neutral fraction, dust absorption, and ionizing escape fraction. The trends we saw in the auto power spectra are still present, but are not as distinct. For example, the slope decreases with neutral fraction as it did in the auto power in Figure~\ref{fig:nf_power_compare}, but not to the same degree. The cross-power slope for each neutral fraction is $P_\mathrm{\MathLya,gal}\propto k^{-1.5}$ for $21\%$, $P_\mathrm{\MathLya,gal}\propto k^{-1.6}$ for $50\%$, and $P_\mathrm{\MathLya,gal}\propto k^{-1.8}$ for $75\%$. We also see when looking at dust absorption, the slope of the power changes the most for the highest value of dust as we observed in Figure~\ref{fig:power_dust_3_panel} and the amplitude of the cross power changes similar to the auto-power. We find a similar yet more significant trend for $f_\mathrm{esc}$, in that the the amplitude and slopes of both the auto and cross-power are decreased by increasing the escape fraction. The cross correlation is sensitive to models of reionization, but we defer full forecasting of these constraints to future work.

\section{Discussion and Conclusion} \label{sec:conclusions}

We have simulated complete \Lya intensity maps during cosmic reionization including radiative transfer in the IGM. These simulations include \Lya photons from galaxies, the edges of ionized bubbles, recombinations in the IGM, and reprocessing of galaxy continuum emission. We also assessed the impact of three varying astrophysical parameters (neutral fraction, dust absorption, and ionizing escape fraction) on the intensity maps and associated power spectra. 

We found the power spectra for our chosen \Lya sources vary in amplitude in response to a variety of factors including the neutral fraction of the IGM, the dust absorption in galaxies, and the escape fraction of ionizing photons. For our fiducial model, the galaxies and continuum are the strongest sources of fluctuations, while the edges and recombinations are negligible. We find that increasing neutral fractions result in higher power for the continuum and recombinations, while higher dust absorption results in a decrease in the galaxy power spectrum. Finally, we find that higher ionizing photon escape fractions result in the edges and recombinations power spectra to outweigh those of the galaxies and continuum. The contribution from galaxies is less than that of recombinations on the largest scales for $f_\mathrm{esc} = 0.3$, and it becomes the least significant source at $f_\mathrm{esc}=0.8$. 

We also found that the slope of the power spectrum can be used to infer the neutral fraction of the IGM, supporting the findings of \cite{Visbal2018}. We note the slope was impacted from varying the dust absorption and ionizing escape fraction, but these we believe these degeneracies have little effect on our ability to constrain the neutral fraction of the IGM. When observing over a range of redshifts, we expect to see less change in the dust absorption and ionizing escape fraction than we would for the neutral fraction, which would allow us to potentially disentangle these degeneracies. The neutral fraction may also be disentangled by using the fact that scattering by neutral gas leads to polarization \citep{2020PhRvD.101h3032M}.

We also investigated the effect of a velocity offset on the \Lya line from galaxies in order to approximate the radiative transfer in the ISM. We found that this velocity offset had little impact on the power spectra. One would require an offset velocity of $v_\mathrm{off} \geq 500$~km/s to significantly alter the power on the smallest scales, a value that is higher than those found in \cite{Blaizot2023,Smith2019,Smith2022}. This indicates that radiative transfer in the IGM is a crucial component for such calculations.

Finally, using \Lya intensity maps from SPHEREx-like observations and a future galaxy redshift survey, we found the cross correlation signal will be detectable. We find a total S/N$\approx4$ for modes $k=0.035-1~$Mpc$^{-1}$. We also take cross correlations for power spectra with different astrophysical parameters. We see similar trends to those we saw previously in the auto-power, but less distinct.

In future work, we will perform a full analysis on the viability of constraining galaxy formation and reionization models with \Lya intensity mapping data using both auto and cross power. This will quantify in detail the degeneracies related to the dust absorption and ionizing escape fraction, as both values impact the slope of the power spectra in qualitatively similar ways.

\acknowledgments
We would like to thank the anonymous referee for their helpful insight which improved the manuscript. The majority of our computations were carried out at the Ohio Supercomputer Center. EV is supported by NSF grant AST-2009309, NASA ATP grant 80NSSC22K0629, and STScI grant JWST-AR-05238. AA is supported by STScI grant JWST-AR-05238.

% Bibliography

%% [A] Recommended: using JHEP.bst file
\bibliographystyle{JHEP}
\bibliography{LyaAlpha.bib}

\providecommand{\href}[2]{#2}\begingroup\raggedright\begin{thebibliography}{10}

\bibitem{Iye2006}
M.~Iye, K.~Ota, N.~Kashikawa, H.~Furusawa, T.~Hashimoto, T.~Hattori et~al., \emph{A galaxy at a redshift z = 6.96}, \href{https://doi.org/10.1038/nature05104}{\emph{Nature} {\bfseries 443} (2006) 186}.

\bibitem{Ono2012}
Y.~{Ono}, M.~{Ouchi}, B.~{Mobasher}, M.~{Dickinson}, K.~{Penner}, K.~{Shimasaku} et~al., \emph{{Spectroscopic Confirmation of Three z-dropout Galaxies at z = 6.844-7.213: Demographics of Ly{\ensuremath{\alpha}} Emission in $z \raisebox{-0.5ex}\textasciitilde 7$ Galaxies}}, \href{https://doi.org/10.1088/0004-637X/744/2/83}{\emph{ApJ} {\bfseries 744} (2012) 83} [\href{https://arxiv.org/abs/1107.3159}{{\ttfamily 1107.3159}}].

\bibitem{Schenker2012}
M.A.~{Schenker}, D.P.~{Stark}, R.S.~{Ellis}, B.E.~{Robertson}, J.S.~{Dunlop}, R.J.~{McLure} et~al., \emph{{Keck Spectroscopy of Faint $3 < z < 8$ Lyman Break Galaxies: Evidence for a Declining Fraction of Emission Line Sources in the Redshift Range $6 < z < 8$}}, \href{https://doi.org/10.1088/0004-637X/744/2/179}{\emph{ApJ} {\bfseries 744} (2012) 179} [\href{https://arxiv.org/abs/1107.1261}{{\ttfamily 1107.1261}}].

\bibitem{Zhu2024}
Y.~{Zhu}, G.D.~{Becker}, S.E.I.~{Bosman}, C.~{Cain}, L.C.~{Keating}, F.~{Nasir} et~al., \emph{{Damping wing-like features in the stacked Ly {\ensuremath{\alpha}} forest: Potential neutral hydrogen islands at $z < 6$}}, \href{https://doi.org/10.1093/mnrasl/slae061}{\emph{MNRAS} {\bfseries 533} (2024) L49} [\href{https://arxiv.org/abs/2405.12275}{{\ttfamily 2405.12275}}].

\bibitem{Planck2016}
{Planck Collaboration}, {Adam, R.}, {Aghanim, N.}, {Ashdown, M.}, {Aumont, J.}, {Baccigalupi, C.} et~al., \emph{Planck intermediate results - xlvii. planck constraints on reionization history}, \href{https://doi.org/10.1051/0004-6361/201628897}{\emph{A\&A} {\bfseries 596} (2016) A108}.

\bibitem{Wise2019}
J.H.~Wise, \emph{Cosmic reionisation}, \href{https://doi.org/10.1080/00107514.2019.1631548}{\emph{Contemporary Physics} {\bfseries 60} (2019) 145} [\href{https://arxiv.org/abs/https://doi.org/10.1080/00107514.2019.1631548}{{\ttfamily https://doi.org/10.1080/00107514.2019.1631548}}].

\bibitem{Robertson2022}
B.E.~{Robertson}, \emph{{Galaxy Formation and Reionization: Key Unknowns and Expected Breakthroughs by the James Webb Space Telescope}}, \href{https://doi.org/10.1146/annurev-astro-120221-044656}{\emph{ARA\&A} {\bfseries 60} (2022) 121} [\href{https://arxiv.org/abs/2110.13160}{{\ttfamily 2110.13160}}].

\bibitem{Visbal2010}
E.~{Visbal} and A.~{Loeb}, \emph{{Measuring the 3D clustering of undetected galaxies through cross correlation of their cumulative flux fluctuations from multiple spectral lines}}, \href{https://doi.org/10.1088/1475-7516/2010/11/016}{\emph{JCAP} {\bfseries 2010} (2010) 016} [\href{https://arxiv.org/abs/1008.3178}{{\ttfamily 1008.3178}}].

\bibitem{Bernal2022}
J.L.~Bernal and E.D.~Kovetz, \emph{Line-intensity mapping: theory review with a focus on star-formation lines}, \href{https://doi.org/10.1007/s00159-022-00143-0}{\emph{The Astronomy and Astrophysics Review} {\bfseries 30} (2022) 5}.

\bibitem{DeBoer2017}
D.R.~DeBoer, A.R.~Parsons, J.E.~Aguirre, P.~Alexander, Z.S.~Ali, A.P.~Beardsley et~al., \emph{Hydrogen epoch of reionization array (hera)}, \href{https://doi.org/10.1088/1538-3873/129/974/045001}{\emph{PASP} {\bfseries 129} (2017) 045001}.

\bibitem{Cleary2022}
K.A.~Cleary, J.~Borowska, P.C.~Breysse, M.~Catha, D.T.~Chung, S.E.~Church et~al., \emph{Comap early science. i. overview}, \href{https://doi.org/10.3847/1538-4357/ac63cc}{\emph{ApJ} {\bfseries 933} (2022) 182}.

\bibitem{Dumitru2019}
S.~{Dumitru}, G.~{Kulkarni}, G.~{Lagache} and M.G.~{Haehnelt}, \emph{{Predictions and sensitivity forecasts for reionization-era [C II] line intensity mapping}}, \href{https://doi.org/10.1093/mnras/stz617}{\emph{MNRAS} {\bfseries 485} (2019) 3486} [\href{https://arxiv.org/abs/1802.04804}{{\ttfamily 1802.04804}}].

\bibitem{Cheng2022}
Y.-T.~Cheng and T.-C.~Chang, \emph{Cosmic near-infrared background tomography with spherex using galaxy cross-correlations}, \href{https://doi.org/10.3847/1538-4357/ac3aee}{\emph{ApJ} {\bfseries 925} (2022) 136}.

\bibitem{Alibay2023}
F.~Alibay, O.V.~Sindiy, P.A.T.~Jansma, C.M.~Reynerson, E.B.~Rice, J.~Rocca et~al., \emph{Spherex preliminary mission overview},  in \emph{2023 IEEE Aerospace Conference}, pp.~1--18, 2023, \href{https://doi.org/10.1109/AERO55745.2023.10115793}{DOI}.

\bibitem{Cooray2019}
A.~{Cooray}, T.-C.~{Chang}, S.~{Unwin}, M.~{Zemcov}, A.~{Coffey}, P.~{Morrissey} et~al., \emph{{Cosmic Dawn Intensity Mapper}},  in \emph{Bulletin of the American Astronomical Society}, vol.~51, p.~23, Sept., 2019, \href{https://doi.org/10.48550/arXiv.1903.03144}{DOI} [\href{https://arxiv.org/abs/1903.03144}{{\ttfamily 1903.03144}}].

\bibitem{Silva2013}
M.B.~{Silva}, M.G.~{Santos}, Y.~{Gong}, A.~{Cooray} and J.~{Bock}, \emph{{Intensity Mapping of Ly{\ensuremath{\alpha}} Emission during the Epoch of Reionization}}, \href{https://doi.org/10.1088/0004-637X/763/2/132}{\emph{ApJ} {\bfseries 763} (2013) 132} [\href{https://arxiv.org/abs/1205.1493}{{\ttfamily 1205.1493}}].

\bibitem{Pullen2014}
A.R.~{Pullen}, O.~{Dor{\'e}} and J.~{Bock}, \emph{{Intensity Mapping across Cosmic Times with the Ly{\ensuremath{\alpha}} Line}}, \href{https://doi.org/10.1088/0004-637X/786/2/111}{\emph{ApJ} {\bfseries 786} (2014) 111} [\href{https://arxiv.org/abs/1309.2295}{{\ttfamily 1309.2295}}].

\bibitem{Heneka2021}
C.~{Heneka} and A.~{Cooray}, \emph{{Optimal survey parameters: Ly {\ensuremath{\alpha}} and H {\ensuremath{\alpha}} intensity mapping for synergy with the 21-cm signal during reionization}}, \href{https://doi.org/10.1093/mnras/stab1842}{\emph{MNRAS} {\bfseries 506} (2021) 1573} [\href{https://arxiv.org/abs/2104.12739}{{\ttfamily 2104.12739}}].

\bibitem{Padmanabhan2024}
H.~{Padmanabhan} and A.~{Loeb}, \emph{{Intensity mapping of Loeb-Rybicki haloes from scattering of galactic Lyman-$\alpha$ emission by the diffuse intergalactic medium before reionization}}, \href{https://doi.org/10.48550/arXiv.2408.16820}{\emph{arXiv e-prints} (2024) arXiv:2408.16820} [\href{https://arxiv.org/abs/2408.16820}{{\ttfamily 2408.16820}}].

\bibitem{Visbal2018}
E.~{Visbal} and M.~{McQuinn}, \emph{{The Impact of Neutral Intergalactic Gas on Ly{\ensuremath{\alpha}} Intensity Mapping during Reionization}}, \href{https://doi.org/10.3847/2041-8213/aad5e6}{\emph{ApJL} {\bfseries 863} (2018) L6} [\href{https://arxiv.org/abs/1807.03370}{{\ttfamily 1807.03370}}].

\bibitem{Planck2018}
{Planck Collaboration}, {Aghanim, N.}, {Akrami, Y.}, {Ashdown, M.}, {Aumont, J.}, {Baccigalupi, C.} et~al., \emph{Planck 2018 results - vi. cosmological parameters}, \href{https://doi.org/10.1051/0004-6361/201833910}{\emph{A\&A} {\bfseries 641} (2020) A6}.

\bibitem{Mesinger2011}
A.~{Mesinger}, S.~{Furlanetto} and R.~{Cen}, \emph{{21CMFAST: a fast, seminumerical simulation of the high-redshift 21-cm signal}}, \href{https://doi.org/10.1111/j.1365-2966.2010.17731.x}{\emph{MNRAS} {\bfseries 411} (2011) 955} [\href{https://arxiv.org/abs/1003.3878}{{\ttfamily 1003.3878}}].

\bibitem{Zahn2007}
O.~{Zahn}, A.~{Lidz}, M.~{McQuinn}, S.~{Dutta}, L.~{Hernquist}, M.~{Zaldarriaga} et~al., \emph{{Simulations and Analytic Calculations of Bubble Growth during Hydrogen Reionization}}, \href{https://doi.org/10.1086/509597}{\emph{ApJ} {\bfseries 654} (2007) 12} [\href{https://arxiv.org/abs/astro-ph/0604177}{{\ttfamily astro-ph/0604177}}].

\bibitem{Sheth1999}
R.K.~Sheth and G.~Tormen, \emph{{Large-scale bias and the peak background split}}, \href{https://doi.org/10.1046/j.1365-8711.1999.02692.x}{\emph{MNRAS} {\bfseries 308} (1999) 119} [\href{https://arxiv.org/abs/https://academic.oup.com/mnras/article-pdf/308/1/119/18409158/308-1-119.pdf}{{\ttfamily https://academic.oup.com/mnras/article-pdf/308/1/119/18409158/308-1-119.pdf}}].

\bibitem{Furlanetto2017}
S.R.~{Furlanetto}, J.~{Mirocha}, R.H.~{Mebane} and G.~{Sun}, \emph{{A minimalist feedback-regulated model for galaxy formation during the epoch of reionization}}, \href{https://doi.org/10.1093/mnras/stx2132}{\emph{MNRAS} {\bfseries 472} (2017) 1576} [\href{https://arxiv.org/abs/1611.01169}{{\ttfamily 1611.01169}}].

\bibitem{yeh2023}
J.Y.-C.~Yeh, A.~Smith, R.~Kannan, E.~Garaldi, M.~Vogelsberger, J.~Borrow et~al., \emph{The thesan project: ionizing escape fractions of reionization-era galaxies}, \href{https://doi.org/10.1093/mnras/stad210}{\emph{MNRAS} {\bfseries 520} (2023) 2757}.

\bibitem{Schaerer2003}
D.~{Schaerer}, \emph{{The transition from Population III to normal galaxies: Lyalpha and He II emission and the ionising properties of high redshift starburst galaxies}}, \href{https://doi.org/10.1051/0004-6361:20021525}{\emph{A\&A} {\bfseries 397} (2003) 527} [\href{https://arxiv.org/abs/astro-ph/0210462}{{\ttfamily astro-ph/0210462}}].

\bibitem{Hayes2011}
M.~{Hayes}, D.~{Schaerer}, G.~{{\"O}stlin}, J.M.~{Mas-Hesse}, H.~{Atek} and D.~{Kunth}, \emph{{On the Redshift Evolution of the Ly{\ensuremath{\alpha}} Escape Fraction and the Dust Content of Galaxies}}, \href{https://doi.org/10.1088/0004-637X/730/1/8}{\emph{ApJ} {\bfseries 730} (2011) 8} [\href{https://arxiv.org/abs/1010.4796}{{\ttfamily 1010.4796}}].

\bibitem{Park2019}
J.~{Park}, A.~{Mesinger}, B.~{Greig} and N.~{Gillet}, \emph{{Inferring the astrophysics of reionization and cosmic dawn from galaxy luminosity functions and the 21-cm signal}}, \href{https://doi.org/10.1093/mnras/stz032}{\emph{MNRAS} {\bfseries 484} (2019) 933} [\href{https://arxiv.org/abs/1809.08995}{{\ttfamily 1809.08995}}].

\bibitem{Yajima2011}
H.~{Yajima}, J.-H.~{Choi} and K.~{Nagamine}, \emph{{Escape fraction of ionizing photons from high-redshift galaxies in cosmological SPH simulations}}, \href{https://doi.org/10.1111/j.1365-2966.2010.17920.x}{\emph{MNRAS} {\bfseries 412} (2011) 411} [\href{https://arxiv.org/abs/1002.3346}{{\ttfamily 1002.3346}}].

\bibitem{Ferrara2013}
A.~{Ferrara} and A.~{Loeb}, \emph{{Escape fraction of the ionizing radiation from starburst galaxies at high redshifts}}, \href{https://doi.org/10.1093/mnras/stt381}{\emph{MNRAS} {\bfseries 431} (2013) 2826} [\href{https://arxiv.org/abs/1209.2123}{{\ttfamily 1209.2123}}].

\bibitem{Xu2016}
H.~{Xu}, J.H.~{Wise}, M.L.~{Norman}, K.~{Ahn} and B.W.~{O'Shea}, \emph{{Galaxy Properties and UV Escape Fractions during the Epoch of Reionization: Results from the Renaissance Simulations}}, \href{https://doi.org/10.3847/1538-4357/833/1/84}{\emph{ApJ} {\bfseries 833} (2016) 84} [\href{https://arxiv.org/abs/1604.07842}{{\ttfamily 1604.07842}}].

\bibitem{Behroozi2015}
P.S.~{Behroozi} and J.~{Silk}, \emph{{A Simple Technique for Predicting High-redshift Galaxy Evolution}}, \href{https://doi.org/10.1088/0004-637X/799/1/32}{\emph{ApJ} {\bfseries 799} (2015) 32} [\href{https://arxiv.org/abs/1404.5299}{{\ttfamily 1404.5299}}].

\bibitem{Mirocha2017}
J.~{Mirocha}, S.R.~{Furlanetto} and G.~{Sun}, \emph{{The global 21-cm signal in the context of the high- z galaxy luminosity function}}, \href{https://doi.org/10.1093/mnras/stw2412}{\emph{MNRAS} {\bfseries 464} (2017) 1365} [\href{https://arxiv.org/abs/1607.00386}{{\ttfamily 1607.00386}}].

\bibitem{Blaizot2023}
J.~{Blaizot}, T.~{Garel}, A.~{Verhamme}, H.~{Katz}, T.~{Kimm}, L.~{Michel-Dansac} et~al., \emph{{Simulating the diversity of shapes of the Lyman-{\ensuremath{\alpha}} line}}, \href{https://doi.org/10.1093/mnras/stad1523}{\emph{MNRAS} {\bfseries 523} (2023) 3749} [\href{https://arxiv.org/abs/2305.10047}{{\ttfamily 2305.10047}}].

\bibitem{Smith2019}
A.~{Smith}, X.~{Ma}, V.~{Bromm}, S.L.~{Finkelstein}, P.F.~{Hopkins}, C.-A.~{Faucher-Gigu{\`e}re} et~al., \emph{{The physics of Lyman {\ensuremath{\alpha}} escape from high-redshift galaxies}}, \href{https://doi.org/10.1093/mnras/sty3483}{\emph{MNRAS} {\bfseries 484} (2019) 39} [\href{https://arxiv.org/abs/1810.08185}{{\ttfamily 1810.08185}}].

\bibitem{Smith2022}
A.~{Smith}, R.~{Kannan}, S.~{Tacchella}, M.~{Vogelsberger}, L.~{Hernquist}, F.~{Marinacci} et~al., \emph{{The physics of Lyman-{\ensuremath{\alpha}} escape from disc-like galaxies}}, \href{https://doi.org/10.1093/mnras/stac2641}{\emph{MNRAS} {\bfseries 517} (2022) 1} [\href{https://arxiv.org/abs/2111.13721}{{\ttfamily 2111.13721}}].

\bibitem{DAloisio2019}
A.~{D'Aloisio}, M.~{McQuinn}, O.~{Maupin}, F.B.~{Davies}, H.~{Trac}, S.~{Fuller} et~al., \emph{{Heating of the Intergalactic Medium by Hydrogen Reionization}}, \href{https://doi.org/10.3847/1538-4357/ab0d83}{\emph{ApJ} {\bfseries 874} (2019) 154} [\href{https://arxiv.org/abs/1807.09282}{{\ttfamily 1807.09282}}].

\bibitem{santos2002}
M.R.~Santos, V.~Bromm and M.~Kamionkowski, \emph{The contribution of the first stars to the cosmic infrared background}, \href{https://doi.org/10.1046/j.1365-8711.2002.05895.x}{\emph{MNRAS} {\bfseries 336} (2002) 1082}.

\bibitem{Wilson2024}
B.~{Wilson}, A.~{D'Aloisio}, G.D.~{Becker}, C.~{Cain} and E.~{Visbal}, \emph{{Imaging reionization's last phases with I-front Lyman-{\ensuremath{\alpha}} emissions}}, \href{https://doi.org/10.1088/1475-7516/2025/01/066}{\emph{JCAP} {\bfseries 2025} (2025) 066} [\href{https://arxiv.org/abs/2406.14625}{{\ttfamily 2406.14625}}].

\bibitem{Iliev2007}
I.T.~{Iliev}, G.~{Mellema}, P.R.~{Shapiro} and U.-L.~{Pen}, \emph{{Self-regulated reionization}}, \href{https://doi.org/10.1111/j.1365-2966.2007.11482.x}{\emph{MNRAS} {\bfseries 376} (2007) 534} [\href{https://arxiv.org/abs/astro-ph/0607517}{{\ttfamily astro-ph/0607517}}].

\bibitem{Mcquinn2007}
M.~McQuinn, A.~Lidz, O.~Zahn, S.~Dutta, L.~Hernquist and M.~Zaldarriaga, \emph{The morphology of {H} ii regions during reionization}, \href{https://doi.org/10.1111/j.1365-2966.2007.11489.x}{\emph{MNRAS} {\bfseries 377} (2007) 1043}.

\bibitem{DAloisio2020}
A.~D’Aloisio, M.~McQuinn, H.~Trac, C.~Cain and A.~Mesinger, \emph{Hydrodynamic response of the intergalactic medium to reionization}, \href{https://doi.org/10.3847/1538-4357/ab9f2f}{\emph{ApJ} {\bfseries 898} (2020) 149}.

\bibitem{Osterbrock2006Book}
D.E.~{Osterbrock} and G.J.~{Ferland}, \emph{{Astrophysics of gaseous nebulae and active galactic nuclei}} (2006).

\bibitem{Haiman1997}
Z.~Haiman, M.J.~Rees and A.~Loeb, \emph{Destruction of molecular hydrogen during cosmological reionization}, \href{https://doi.org/10.1086/303647}{\emph{ApJ} {\bfseries 476} (1997) 458}.

\bibitem{Haiman2000}
Z.~Haiman, T.~Abel and M.J.~Rees, \emph{The radiative feedback of the first cosmological objects}, \href{https://doi.org/10.1086/308723}{\emph{ApJ} {\bfseries 534} (2000) 11}.

\bibitem{Machacek2001}
M.E.~Machacek, G.L.~Bryan and T.~Abel, \emph{Simulations of pregalactic structure formation with radiative feedback}, \href{https://doi.org/10.1086/319014}{\emph{ApJ} {\bfseries 548} (2001) 509}.

\bibitem{Wise2007}
J.H.~Wise and T.~Abel, \emph{Suppression of h2 cooling in the ultraviolet background}, \href{https://doi.org/10.1086/522876}{\emph{ApJ} {\bfseries 671} (2007) 1559}.

\bibitem{OShea2015}
B.W.~O’Shea, J.H.~Wise, H.~Xu and M.L.~Norman, \emph{Probing the ultraviolet luminosity function of the earliest galaxies with the renaissance simulations}, \href{https://doi.org/10.1088/2041-8205/807/1/L12}{\emph{ApJ} {\bfseries 807} (2015) L12}.

\bibitem{Ahn2009}
K.~{Ahn}, P.R.~{Shapiro}, I.T.~{Iliev}, G.~{Mellema} and U.-L.~{Pen}, \emph{{The Inhomogeneous Background Of H$_{2}$-Dissociating Radiation During Cosmic Reionization}}, \href{https://doi.org/10.1088/0004-637X/695/2/1430}{\emph{ApJ} {\bfseries 695} (2009) 1430} [\href{https://arxiv.org/abs/0807.2254}{{\ttfamily 0807.2254}}].

\bibitem{Visbal2014}
E.~Visbal, Z.~Haiman, B.~Terrazas, G.L.~Bryan and R.~Barkana, \emph{{High-redshift star formation in a time-dependent Lyman–Werner background}}, \href{https://doi.org/10.1093/mnras/stu1710}{\emph{MNRAS} {\bfseries 445} (2014) 107} [\href{https://arxiv.org/abs/https://academic.oup.com/mnras/article-pdf/445/1/107/18472155/stu1710.pdf}{{\ttfamily https://academic.oup.com/mnras/article-pdf/445/1/107/18472155/stu1710.pdf}}].

\bibitem{Barkana2005}
R.~{Barkana} and A.~{Loeb}, \emph{{Detecting the Earliest Galaxies through Two New Sources of 21 Centimeter Fluctuations}}, \href{https://doi.org/10.1086/429954}{\emph{ApJ} {\bfseries 626} (2005) 1} [\href{https://arxiv.org/abs/astro-ph/0410129}{{\ttfamily astro-ph/0410129}}].

\bibitem{Dunlop2012}
J.S.~{Dunlop}, R.J.~{McLure}, B.E.~{Robertson}, R.S.~{Ellis}, D.P.~{Stark}, M.~{Cirasuolo} et~al., \emph{{A critical analysis of the ultraviolet continuum slopes ({\ensuremath{\beta}}) of high-redshift galaxies: no evidence (yet) for extreme stellar populations at $z > 6$}}, \href{https://doi.org/10.1111/j.1365-2966.2011.20102.x}{\emph{MNRAS} {\bfseries 420} (2012) 901} [\href{https://arxiv.org/abs/1102.5005}{{\ttfamily 1102.5005}}].

\bibitem{cullen2023}
F.~Cullen, R.J.~McLure, D.J.~McLeod, J.S.~Dunlop, C.T.~Donnan, A.C.~Carnall et~al., \emph{The ultraviolet continuum slopes ($\beta$) of galaxies at $z \simeq 8-16$ from jwst and ground-based near-infrared imaging}, \href{https://doi.org/10.1093/mnras/stad073}{\emph{MNRAS} {\bfseries 520} (2023) 14}.

\bibitem{Yang2020}
J.~{Yang}, F.~{Wang}, X.~{Fan}, J.F.~{Hennawi}, F.B.~{Davies}, M.~{Yue} et~al., \emph{{Measurements of the z {\ensuremath{\sim}} 6 Intergalactic Medium Optical Depth and Transmission Spikes Using a New $z > 6.3$ Quasar Sample}}, \href{https://doi.org/10.3847/1538-4357/abbc1b}{\emph{ApJ} {\bfseries 904} (2020) 26} [\href{https://arxiv.org/abs/2009.13544}{{\ttfamily 2009.13544}}].

\bibitem{Hirata2006}
C.M.~{Hirata}, \emph{{Wouthuysen-Field coupling strength and application to high-redshift 21-cm radiation}}, \href{https://doi.org/10.1111/j.1365-2966.2005.09949.x}{\emph{MNRAS} {\bfseries 367} (2006) 259} [\href{https://arxiv.org/abs/astro-ph/0507102}{{\ttfamily astro-ph/0507102}}].

\bibitem{Faucher2010}
C.-A.~{Faucher-Gigu{\`e}re}, D.~{Kere{\v{s}}}, M.~{Dijkstra}, L.~{Hernquist} and M.~{Zaldarriaga}, \emph{{Ly{\ensuremath{\alpha}} Cooling Emission from Galaxy Formation}}, \href{https://doi.org/10.1088/0004-637X/725/1/633}{\emph{ApJ} {\bfseries 725} (2010) 633} [\href{https://arxiv.org/abs/1005.3041}{{\ttfamily 1005.3041}}].

\bibitem{Zheng2002}
Z.~{Zheng} and J.~{Miralda-Escud{\'e}}, \emph{{Monte Carlo Simulation of Ly{\ensuremath{\alpha}} Scattering and Application to Damped Ly{\ensuremath{\alpha}} Systems}}, \href{https://doi.org/10.1086/342400}{\emph{ApJ} {\bfseries 578} (2002) 33} [\href{https://arxiv.org/abs/astro-ph/0203287}{{\ttfamily astro-ph/0203287}}].

\bibitem{Cantalupo2005}
S.~{Cantalupo}, C.~{Porciani}, S.J.~{Lilly} and F.~{Miniati}, \emph{{Fluorescent Ly{\ensuremath{\alpha}} Emission from the High-Redshift Intergalactic Medium}}, \href{https://doi.org/10.1086/430758}{\emph{ApJ} {\bfseries 628} (2005) 61} [\href{https://arxiv.org/abs/astro-ph/0504015}{{\ttfamily astro-ph/0504015}}].

\bibitem{Dijkstra2006}
M.~{Dijkstra}, Z.~{Haiman} and M.~{Spaans}, \emph{{Ly{\ensuremath{\alpha}} Radiation from Collapsing Protogalaxies. I. Characteristics of the Emergent Spectrum}}, \href{https://doi.org/10.1086/506243}{\emph{ApJ} {\bfseries 649} (2006) 14} [\href{https://arxiv.org/abs/astro-ph/0510407}{{\ttfamily astro-ph/0510407}}].

\bibitem{Laursen2007}
P.~{Laursen} and J.~{Sommer-Larsen}, \emph{{Ly{\ensuremath{\alpha}} Resonant Scattering in Young Galaxies: Predictions from Cosmological Simulations}}, \href{https://doi.org/10.1086/513191}{\emph{ApJL} {\bfseries 657} (2007) L69} [\href{https://arxiv.org/abs/astro-ph/0610761}{{\ttfamily astro-ph/0610761}}].

\bibitem{Ouchi2018}
M.~{Ouchi}, Y.~{Harikane}, T.~{Shibuya}, K.~{Shimasaku}, Y.~{Taniguchi}, A.~{Konno} et~al., \emph{{Systematic Identification of LAEs for Visible Exploration and Reionization Research Using Subaru HSC (SILVERRUSH). I. Program strategy and clustering properties of {\ensuremath{\sim}}2000 Ly{\ensuremath{\alpha}} emitters at z = 6-7 over the 0.3-0.5 Gpc$^{2}$ survey area}}, \href{https://doi.org/10.1093/pasj/psx074}{\emph{PASJ} {\bfseries 70} (2018) S13} [\href{https://arxiv.org/abs/1704.07455}{{\ttfamily 1704.07455}}].

\bibitem{Higuchi2019}
R.~Higuchi, M.~Ouchi, Y.~Ono, T.~Shibuya, J.~Toshikawa, Y.~Harikane et~al., \emph{Silverrush. vii. subaru/hsc identifications of protocluster candidates at $z\sim6–7$: Implications for cosmic reionization}, \href{https://doi.org/10.3847/1538-4357/ab2192}{\emph{ApJ} {\bfseries 879} (2019) 28}.

\bibitem{Comaschi2016}
P.~{Comaschi} and A.~{Ferrara}, \emph{{Probing high-redshift galaxies with Ly{\ensuremath{\alpha}} intensity mapping}}, \href{https://doi.org/10.1093/mnras/stv2339}{\emph{MNRAS} {\bfseries 455} (2016) 725} [\href{https://arxiv.org/abs/1506.08838}{{\ttfamily 1506.08838}}].

\bibitem{2020PhRvD.101h3032M}
L.~{Mas-Ribas} and T.-C.~{Chang}, \emph{{Lyman-{\ensuremath{\alpha}} polarization intensity mapping}}, \href{https://doi.org/10.1103/PhysRevD.101.083032}{\emph{PhRvD} {\bfseries 101} (2020) 083032} [\href{https://arxiv.org/abs/2002.04107}{{\ttfamily 2002.04107}}].

\end{thebibliography}\endgroup

%% or
%% [B] Manual formatting (see below)
%% (i) We suggest to always provide author, title and journal data or doi:
%% in short all the informations that clearly identify a document.
%% (ii) please avoid comments such as "For a review'', "For some examples",
%% "and references therein" or move them in the text. In general, please leave only references in the bibliography and move all
%% accessory text in footnotes.
%% (iii) Also, please have only one work for each \bibitem.

%\begin{thebibliography}{99}

%\bibitem{a}
%Author,
%\emph{Title},
%\emph{J. Abbrev.} {\bf vol} (year) pg.
%\bibitem{b}
%Author,
%\emph{Title},
%arxiv:1234.5678.

%\bibitem{c}
%Author,
%\emph{Title},
%Publisher (year).

%\end{thebibliography}
\end{document}